\documentclass[%
singlecolumn,
amsmath,amssymb,
aps,
notitlepage
]{revtex4-2}

\usepackage{natbib}
\usepackage{amsmath,amssymb}

\usepackage{graphicx}  
\usepackage{dcolumn}   
\usepackage{bm}        
\usepackage{xcolor}
\usepackage{ulem}
\usepackage{epstopdf}
\usepackage{hyperref}
\usepackage{color}

\usepackage{cancel}

\usepackage{subfigmat} 
\usepackage{threeparttable} 

\DeclareMathOperator{\sech}{sech}

\begin{document}

\preprint{APS/123-QED}

\title{Enduring two-dimensional perturbations with significant non-modal growth}

\author{Sharath~Jose}
\affiliation{International Centre for Theoretical Sciences, Bengaluru, 560089, India.}
\email{sharath872000@gmail.com}

\date{}

\begin{abstract}
  Laminar shear flows can display large non-modal perturbation growth, often through the lift-up mechansm, and can undergo subcritical transition to turbulence. The process is three-dimensional. Two-dimensional (2D) spanwise-independent perturbations are often considered less important as they typically undergo modest levels of transient growth and are short-lived. Strikingly, we show the existence of 2D non-modal perturbations that get amplified significantly and survive for long periods of time. Two-layer and three-layer viscosity stratified plane shear flows are taken to be the mean states. We show that while the two-layer flow is always modally stable, the three-layer flow supports exponential growing instabilities only when the middle layer is the least viscous. The non-modal stability analysis is performed only for the modally stable configurations of these flows. At later times, the non-modal perturbations feature strongly confined vortical structures near the interface in the two-layer flow. For the three-layer flow, similar observations are noted when all the three layers have different shear rates with the vortices prominently seen in the vicinity of the interface between the least viscous and middle layers. For the three-layer flow configuration with the outer layers having equal shear rates, the perturbation structure shows symmetry about the middle layer and evolves such that the Orr mechanism can repeatedly occur in a regenerative manner resulting in the perturbation energy evolving in a markedly non-monotonic fashion. When these same perturbations are introduced in a uniform plane shear flow, the extent of non-modal transient growth is shown to be significantly smaller. 
\end{abstract}

\maketitle

\section{Introduction}
\label{sec:intro}
For examining stability of plane parallel shear flows, non-modal analysis is now firmly established as an invaluable complement to the more traditional modal analysis \citep{Schmid_Henningson_2001book,Schmid_2007ARFM,Schmid_Brandt_2014AMR}. Transient algebraic amplification of perturbations can be observed even when the modal theory predicts all the eigenmodes to decay exponentially in time as the linearised stability operator is non-normal \citep{Butler_Farrell_1992PFA,Trefethen_etal_1993Science}. The transient growth might be significant enough such that nonlinear effects become important enough so as to trigger transition in flows \cite{Bottin_etal_1998PF,Elofsson_etal_1999PF,Andersson_etal_2001JFM}. Even for modally unstable configurations, the sub-critical mechanisms can either act to substantially increase the amplitude of the unstable mode or dominate the transition dynamics outright \cite{Schrader_etal_11JFM,Lucas_etal_15EJMBF,Jose_etal_2017PRF}. Furthermore, unstable modes can be shown to arise as a consequence of non-normality of the underlying linearised stability operator \citep{Jose_Govindarajan_2020PRSA}. 

The two well-known mechanisms behind non-modal behaviour are the three-dimensional (3D) lift-up effect \citep{Landahl_1980JFM,Brandt_2014EJMBF} and the two-dimensional (2D) Orr mechanism \citep{Orr_1907PRIAA_alt}. In canonical plane shear flows, the lift-up effect is more dominant with regard to energy amplification. The lift-up effect arises when a streamwise vortical perturbation evolves to yield streamwise streaks. Evoking a displaced particle argument, the streamwise vortex is said to move fluid elements from a lower velocity region to one having a higher velocity and vice versa. Dynamics involving streamwise vortices and streaks are known to be extremely prevalent in self-sustaining cycle of shear flow turbulence \citep{Waleffe_1997PF,Jimenez_Pinelli_1999JFM,Panton_2001PAS,Schoppa_Hussain_2002JFM,Lozano-Duran_etal_2021JFM}.

In contrast, the Orr mechanism is a comparatively short-lived phenomenon with the associated energy growth being far smaller than what is observed during the lift-up effect. Nonetheless, the Orr mechanism has been credited to be important in bursting events in wall-bounded shear turbulence \citep{Jimenez_2013PF,Jimenez_2018JFM}. Furthermore, wave packets in the region downstream of the potential core in turbulent jets have been shown to undergo amplification due to the Orr mechanism \citep{Tissot_etal_2017PRF,Pickering_etal_2020JFM}. The scenarios in which the Orr mechanism acts in conjunction with the lift-up effect are relevant for finding non-modal optimal perturbations in nonparallel flows \citep{Hack_Moin_2017JFM} and in the transition of parallel flow due to oblique waves \citep{Reddy_etal_1998JFM,Jiao_etal_2021PRF}. 

The focus of the present study will be on the dynamics of 2D non-modal perturbations. In this regard, it is useful to highlight what type of perturbations can undergo energy amplification. With $\psi$ representing the streamfunction of the perturbation to the streamwise mean velocity $U(y)$, the evolution equation for the volume averaged perturbation energy $E$ is:
\begin{align}
  \frac{\textrm{d} E}{\textrm{d}t} = \int_{V} \textrm{d}V\,\frac{\partial \psi}{\partial x}\frac{\partial \psi}{\partial y}U' = \int_{V} \textrm{d}V\,\frac{\partial \psi/\partial x}{\partial \psi/\partial y}\left(\frac{\partial \psi}{\partial y}\right)^2U' = -\int_{V} \textrm{d}V\,\left(\frac{\partial y}{\partial x}\right)_{\psi}\left(\frac{\partial \psi}{\partial y}\right)^2U'. \label{eq:Eevol_1}
\end{align}
In the above, $U' = \textrm{d}U/\textrm{d}y$. From equation \eqref{eq:Eevol_1}, it is apparent that the contribution to rate of change of $E$ is positive in those regions where the wavefunction front is inclined against the mean shear. In the case of uniform shear flow, the mean flow tilts the front until it becomes aligned with the mean shear; this is precisely the physical interpretation of the Orr mechanism \citep{Roy_Govindarajan_2010BookChap}.

We wish to explore some non-modal scenarios for perturbation amplification where the mean shear is no longer uniform. The simplest such departure from the plane shear flow is one where the mean velocity profile has effectively piecewise constant shear rates. To this end, the base flows considered in this work are stratified shear flows with density matched fluid layers with distinguishable viscosities.
While viscosity stratification can serve to stabilise or destabilise the flow \citep{Govindarajan_Sahu_2014ARFM}, its role in this study is limited to defining the mean flow.
As the lift-up effect and the Orr mechanism can be described without invoking viscous effects \citep{Ellingsen_Palm_1975PF,Roy_Subramanian_2014JFM}, the ensuing discussion of the perturbation dynamics in the current study is also inviscid. 

The outline of the paper is as follows. We describe the base flows considered for this study in section \ref{sec:base_flow}. Two-layer and three-layer plane shear flows are considered. To set the stage for non-modal analysis, the modal stability characteristics of these flows are first presented in section \ref{sec:mod_stab}. We show that while the two-layer flow is always modally stable, three-flow configurations with the middle layer being the least viscous are unstable. The characteristics of non-modal perturbations and their evolution are then described for both two-layer and three-layer flows in section \ref{sec:nonmod_stab}. Strikingly, in contrast to the unstratified system, there exist long-lived 2D non-modal perturbations that exhibit significant transient growth. One principal finding in the current study is the emergence of a regenerative Orr mechanism that manifests in the form of the perturbations not undergoing rapid decay after the transient growth phase. Finally, the results are summarised and potential extensions to this work are touched upon in section \ref{sec:conc}.

\section{Base flow configurations}
\label{sec:base_flow}
\begin{figure}
 \centering
 \includegraphics[width=\textwidth]{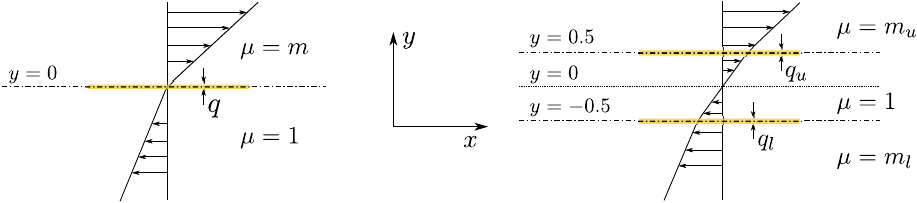}
 \caption{(a) Two-layer viscosity stratified flow. (b) Three-layer viscosity stratified flow. The viscosities of the different layers are distinct. The locations of the interfaces are shown using dash-dot lines. The fluid viscosity varies continuously across the shaded regions.}
 \label{fig:base_configs}
\end{figure}

Layered flows have been successfully employed to shed light on the fundamental nature of several instabilities \citep{Craik_1985book,Smyth_Carpenter_2019book}. The impetus for understanding viscosity stratification effects on stability also stemmed from one such study by \citet{Yih_1967JFM}. The base shear flow configurations along the streamwise coordinate $x$ with different viscosity stratifications utilised for this study are shown in figure \ref{fig:base_configs}. In the two-layer flow, the layers are assumed to be sufficiently deep such that wall effects can be neglected as they would be far from the interface. Similarly, for the three-layer flow, the depths of the lower and upper layers are taken to be much larger than that of the middle layer. While the fluid viscosity in each layer is distinct, we consider the fluids to be weakly miscible such that the viscosity changes smoothly (and monotonically) from one layer to the next through a small mixed region. The fluids in the different layers are density matched, and surface tension effects are neglected. 

As there is no imposed pressure gradient, the base state velocity $U$ is then obtained by solving:
\begin{align}
  \frac{\textrm{d}}{\textrm{d}y}\left(\mu\frac{\textrm{d}U}{\textrm{d}y}\right) = 0. \label{eq:Eq_NSE}
\end{align}
In the above, $\mu$ is the mean viscosity profile whose precise forms for two-layer and three-layer shear flows are given in the subsections below. While the perturbation analysis used in this study is inviscid, the different layers of the mean flows will continue to be identified in terms of its mean viscosity. 

\subsection{Two-layer flow}
For the two-layer flow (see figure \ref{fig:base_configs} (a)), we align our coordinate system such that $y = 0$ is coincident with the interface. The spatial coordinates are non-dimensionalised such that we have $U(0.5) - U(-0.5) = 1$ for all configurations. The non-dimensional mean viscosity profile is defined to be:
\begin{align}
  \mu(y) = 1 + \frac{(m - 1)}{2}\left[1 + \tanh\left(\frac{y}{q}\right)\right]. \label{eq:mu_diff_2LB}
\end{align}
The thickness of the mixed region $q$ is specified to be small ($q \ll 1$). Away from the interface, the fluid viscosities in the lower and upper layers are 1 and $m$ respectively. 

\subsection{Three-layer flow}
For this system (see figure \ref{fig:base_configs} (b)), we choose the reference scales to be defined based on quantities pertaining to the middle layer. We also take the mid-point of the central layer to lie at $y = 0$. The non-dimensional fluid viscosity profile is given by: 
\begin{align}
  \mu(y) = m_l + \frac{\left(1 - m_l\right)}{2}\left[1 + \tanh \left(\frac{y + 0.5}{q_l}\right)\right] + \frac{\left(m_u - 1\right)}{2}\left[1 + \tanh \left(\frac{y - 0.5}{q_u}\right)\right]. \label{eq:mu_diff_3LB}
\end{align}
As in the two-layer flow, the thickness of the mixed regions are specified to be small ($q_l,q_u \ll 1$). For a fair comparison between flows for different sets of parameters, we fix $U\left(0.5\right) - U\left(-0.5\right) = 1$.

\section{Modal stability characteristics}
\label{sec:mod_stab}
We discuss in this section whether these flow configurations can support exponentially growing modal instabilities in the inviscid setting. The perturbations are taken to be of the form $f(x,y,t) = \tilde{f}(y)\exp(i\alpha(x - ct))$, where $\alpha$ is the streamwise wavenumber and $c$ is the complex phase speed. The Rayleigh equation \citep{Drazin_Reid_2004book,Schmid_Henningson_2001book} then determines the inviscid linear modal stability of the base flow $U$. It is given by:
\begin{align}
  (U-c)\left(\frac{\textrm{d}^2}{\textrm{d}y^2} - \alpha^2\right)\tilde{v} - U''\tilde{v} = 0. \label{eq:Ray_mod}
\end{align}
In the above, $\tilde{v}$ is the normal component of the perturbation velocity. If there exists at least one eigenvalue $c$ with a positive imaginary component, then the flow is unstable. From Rayleigh's inflection point theorem \citep{Drazin_Reid_2004book}, for an exponentially growing instability to exist, the second derivative of $U$ should change sign somewhere in the domain. For base flows with continuous derivatives, the eigenvalue problem given by equation \eqref{eq:Ray_mod} is solved numerically (see appendix \ref{app:num_meth_eval}). We also appeal to Howard's semicircle theorem \citep{Howard_1961JFM,Drazin_Reid_2004book} to filter out numerically spurious eigenvalues. For further insight, we will also refer to the analytically obtained dispersion relations of the sharp interface counterparts of these flows (see Appendix \ref{app:PL_mod}). 

\subsection{Two-layer flow}
When there exists a mixed region across which the viscosity varies smoothly, $U''$ is not always 0; however, it may be quite small. When the fluid viscosity $\mu$ is given by equation (\ref{eq:mu_diff_2LB}), we have:
\begin{align}
 U'' = -\frac{\mu'}{\mu^2}, \textrm{ with } \mu' = \frac{(m - 1)}{2\delta}\sech^2\left[\frac{y}{q}\right].
\end{align}
From the above, it is clear that $U''$ does not change sign. Therefore the two-layer flows do not support inviscid modal instabilities. As long as $\mu$ varies monotonically in the mixed layer, this conclusion holds true.

\subsection{Three-layer flow}
\label{ssec:3L_mod}
\begin{figure}
  \centering
  \includegraphics[width=\textwidth]{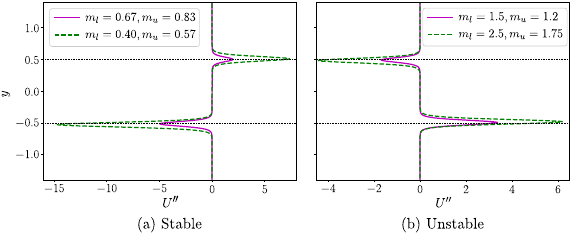}
  \caption{The profiles of $U''$ for different base flow configurations. Although, both sets of configurations satisfy Rayleigh's criterion, only those on the right satisfy Fj\o{}rtoft's criterion. The interface locations are indicated by the horizontal thin dotted lines.}
  \label{fig:3LB_DI_Upp}
\end{figure}

For the three-layer flow, the second derivative $U''$ is given by:
\begin{align}
 U'' = -\frac{\mu'}{\mu^2}, \textrm{ with } \mu' = \frac{(1 - m_l)}{2q_l}\sech^2\left[\frac{y+0.5}{q_l}\right] + \frac{(m_u - 1)}{2q_u}\sech^2\left[\frac{y-0.5}{q_u}\right].
\end{align}
Let us first consider the case where the fluid in the middle layer is neither the most or the least viscous of the three layers, which implies one of either $m_l < 1 < m_u$ or $m_u < 1 < m_l$. It is then evident that $U''$ does not change sign anywhere, and hence inviscid modal instabilities can not exist as Rayleigh's inflection point criterion is not satisfied.

On the other hand, when the outer layers are either more or less viscous than the middle layer, it is seen that $U''$ does end up changing sign in the domain (see figure \ref{fig:3LB_DI_Upp}). In such cases, we find that Fj\o{}rtoft's criterion \citep{Drazin_Reid_2004book}, which additionally requires the inflection point to correspond to a maxima of the square of the mean vorticity for monotonic base flows \citep{Huerre_Rossi_1998BookChap}, has to be satisfied for the existence of modal instabilities. When the middle layer is the most viscous of the three, this necessary condition is not satisfied. On approximating the mean flow by a piecewise linear continuous profile with abrupt jumps in the mean shear rate at the interfaces, we can examine analytically the stability characteristics. Despite considering this extreme condition with a discontinuity in the velocity gradient at the interface, modal stability is always assured when the shear rate in the middle layer is the least (see appendix \ref{app:PL_mod} for details). 

\begin{figure}
  \centering
  \includegraphics[width=0.45\textwidth]{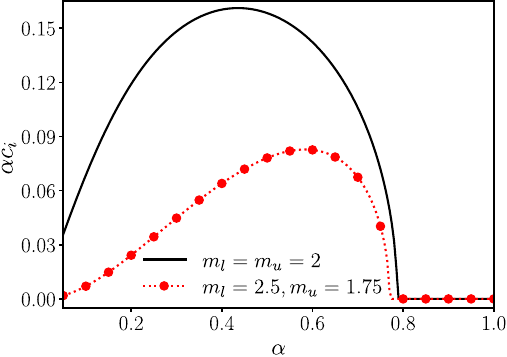}
  \caption{Growth rates ($\alpha c_i$) of the least stable mode as a function of streamwise wavenumber ($\alpha$).}
  \label{fig:3L_gr_v_alp_inv}
\end{figure}

We now focus on cases where the middle layer is less viscous than the outer layers. Figure \ref{fig:3L_gr_v_alp_inv} show the growth rates of the least stable modes for two different flow configurations as a function of the perturbation wavenumber. For modes with short wavelengths ($\alpha > 1$), all the eigenmodes are found to be neutrally stable. There exist unstable modes with $\alpha \lessapprox 0.8$. The growth rates shown here are also seen to be close to those predicted by the analytical analysis for the piecewise linear continuous base flow (not shown). The wavenumber corresponding to the most unstable mode is dependent on the configuration considered. The growth rate of the unstable mode is seen to become lower as the wavenumber reduces. 

\begin{figure}
  \centering
  \includegraphics[width=\textwidth]{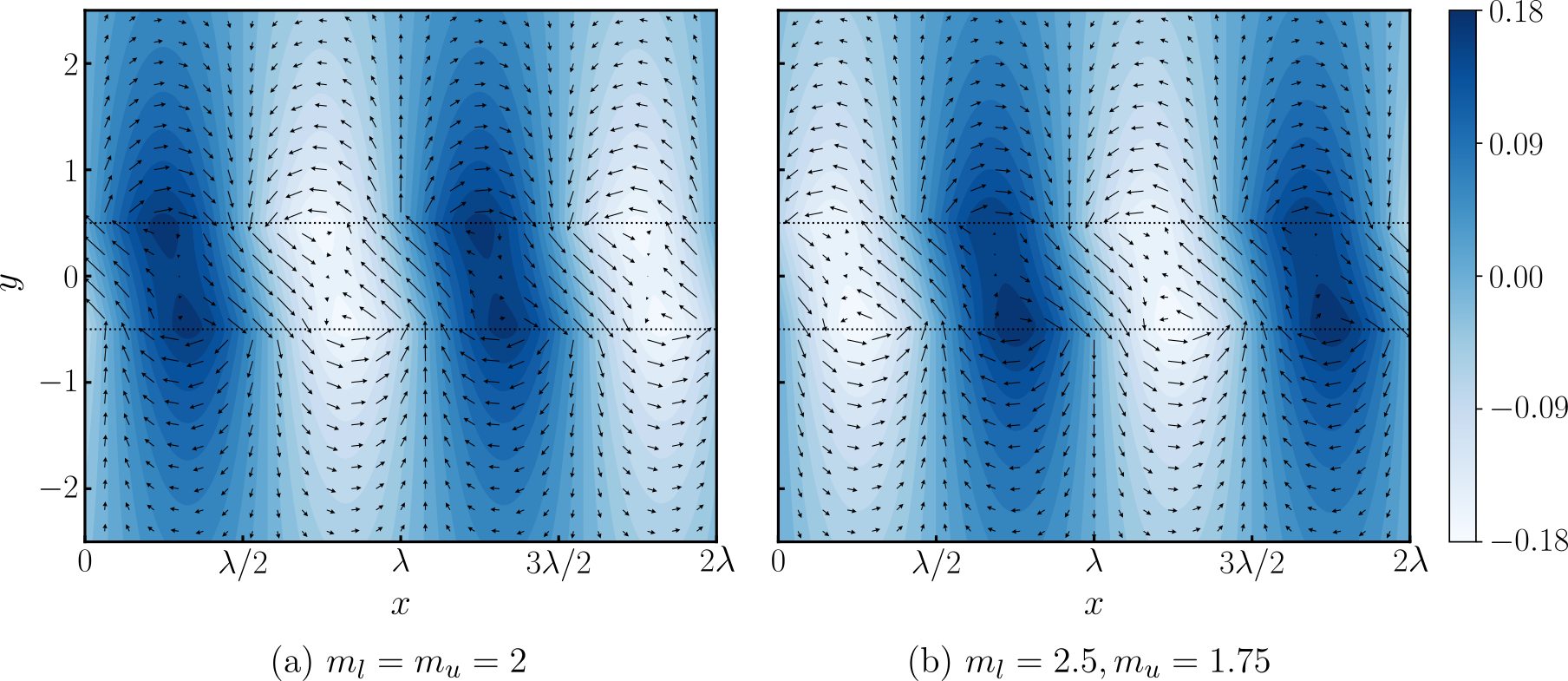}
  \caption{Exponentially unstable modes (inviscid) with $\alpha = 0.5$. The colour and the arrows denote the perturbation streamfunction and the perturbation velocity vector respectively; the interface locations are indicated by the horizontal thin dotted lines.}
  \label{fig:3L_EFs}
\end{figure}

When examining the structure of the unstable modes, it is seen that associated streamfunction is significantly more prominent in the vicinity of the middle layer (e.g., see figure \ref{fig:3L_EFs}). In fact, the regions closest to the interfaces are where most of the production of perturbation kinetic energy occurs. Away from the middle layer, the streamfunction decays quickly. Asymmetries arise when the upper and lower layers have different viscosities. In this case, the streamfunction is stronger near the interface where the change in background shear rate is more drastic. When the shear rates in the upper and lower layers are equal, the phase speed of the unstable mode equals the mean speed at the mid point of the central layer. In the reference frame considered here, these unstable modes are stationary. When the viscosities in the upper and lower layers are unequal, then the unstable mode has a positive (negative) phase speed when the shear rate in the upper (lower) layer is higher. The structure of the unstable modes suggest that the underlying mechanism can be explained in terms of interaction of vorticity waves at the two interfaces \citep{Carpenter_etal_2012AMR}. 

\subsection{Additional comments}

Let us consider the stability analysis of piecewise linear base flows (see appendix \ref{app:PL_mod}). In the plane shear flow of a single layer, \citep{Case_1960PF} showed that no modal solutions exist. In the two-layer flow, the dispersion relation is a linear equation in $c$ with real coefficients as there is only one interface. For the three-layer flow, which has two interfaces, the dispersion relation is now given by a quadratic equation with real coefficients. Mathematically, this allows for the phase speed $c$ to take on complex conjugate values for certain parameter ranges. This idea can be extended to for shear flows with arbitrary number of fluid layers. Therefore, plane shear flows of three or more fluid layers, where the shear rate in each layer is different from those of its neighbours, can potentially be susceptible to exponentially growing instabilities.

The three-layer shear flow considered here has a vorticity profile similar to that of the piecewise linear shear layer (e.g., see \cite{Schmid_Henningson_2001book}). The crucial difference for the flow considered here is that the vorticity is non-zero outside of the central layer. If one were to analyse the smoothened shear layer profile such that all derivatives of the mean flow are continuous, it can be verified that Fj\o{}rtoft's criterion is always satisfied by the unstable configurations.
Earlier work have shown the three-layer immiscible plane Couette flow \citep{Li_1969PF} and the three-layer immiscible flow down an incline \citep{Weinstein_Kurz_1991PFA} to support long-wave instabilities; note that these studies account for the effect of surface tension at the interface and viscous terms. The unstable mean flow configuration for those systems, as in the current study, requires the central layer to be the least viscous of the three. 

\section{Non-modal stability analysis}
\label{sec:nonmod_stab}
As the two-layer flow is always stable, it is straightforward to proceed with the non-modal analysis. On the other hand, for the three-layer flow, the exponential growth rates associated with the instabilities are considerably high (see figure \ref{fig:3L_gr_v_alp_inv}). It is apparent that the exponential instabilities will end up dominating the perturbation dynamics. This is unlike scenarios where the modal instability having a small exponential growth allows for non-modal mechanisms to significantly increase the perturbation amplitude \citep{Jose_etal_2017PRF}. Therefore, the non-modal analysis is restricted to the modally stable configurations for three-layer flows. 

\subsection{Problem formulation and other definitions}
The perturbations are now considered to be of the form $f(x,y,t) = \hat{f}(y,t)\exp\left(i\alpha x\right)$. When these perturbations can not be described in terms of any single eigenmode of the linearised operator, their structure changes over the course of its evolution. The governing equation for the linear perturbation in the absence of viscous effects is:
\begin{align}
  &\mathcal{F}\hat{v} \equiv \partial_t\left(\partial_y^2 - \alpha^2\right)\hat{v} + i\alpha U(\partial_y^2 - \alpha^2)\hat{v} - i\alpha U''\hat{v} = 0. \label{eq:Ray_ft}
\end{align}
$\hat{v}$ is the normal component of the perturbation velocity.

We use the direct-adjoint looping procedure \citep{Luchini_Bottaro_2014ARFM} to find an initial condition that maximises an objective functional (see Appendix \ref{app:DAL} for details). The objective functional for this study is the gain in perturbation energy at $t = T$ for a given set of base flow and perturbation parameters. In terms of $\hat{v}$, the perturbation energy is given by:
\begin{align}
  &E(t) = \frac{1}{4\alpha^2}\int^{\Sigma/2}_{-\Sigma/2} \textrm{d}y\;\partial_y\hat{v}\partial_y\hat{v}^* + \alpha^2\hat{v}\hat{v}^* =  \frac{1}{8\alpha^2}\int^{\Sigma/2}_{-\Sigma/2} \textrm{d}y\; \left(\hat{v}^*\mathcal{M}\hat{v} + \hat{v}\mathcal{M}\hat{v}^*\right). \label{eq:E}
\end{align}
In the above, $\hat{v}^*$ represents the complex conjugate of $\hat{v}$ and $\mathcal{M} = \left(\alpha^2 - \partial_y^2\right)$. Throughout this work, the non-modal perturbations are normalised such that $E(0) = 1$.

For analysing the non-modal perturbations, we also consider the volume-averaged rate of change of perturbation energy, or simply, production:
\begin{align}
  &\frac{\textrm{d}E}{\textrm{d}t} = -\frac{1}{4}\int^{\Sigma/2}_{-\Sigma/2} \textrm{d}y\;U'\left(\hat{v}\hat{u}^* + \hat{u}\hat{v}^*\right).\label{eq:Prod} 
\end{align}
In the above, $\hat{u} = i\partial_y\hat{v}/\alpha$.
For the form of perturbations considered here, note that equations \eqref{eq:Eevol_1} and \eqref{eq:Prod} are equivalent.

Non-uniform fluid properties often end up breaking symmetry of the perturbation structures \citep{Jose_etal_2020IJHFF,Thakur_etal_2021JFM}. The multi-layer plane shear flows offer us a natural framework for analysing the perturbation evolution in different regions. In this regard, the distribution of perturbation energy and its production in different layers are useful quantitative measures. While evaluating these expressions for different layers, the limits of the integrals in equations \eqref{eq:E} and \eqref{eq:Prod} are suitably modified.

We also include some discussion on the structure of the non-modal perturbation at different times during its evolution in the following subsections. Let $l = l(t)$ be the largest magnitude of perturbation velocity at any time. Subsequently, provided the perturbation has evolved for a sufficiently long period, we define $L = \max_{\forall t} l$. For each snapshot in time, the velocity is scaled by $l$ so that the orientation of the vector field comes out clearly. In order to give the true magnitude of the velocity, $l$ will be specified as a fraction of $L$.

For all the results to follow, the computational domain is $ y \in [-0.5\Sigma,0.5\Sigma]$ with $\Sigma = 11$. The upper and lower layers are specified to have equal depths for both the mean flows considered. These depths are large enough such that there is no undue influence of the walls on the dynamics. We also note that while the choice of the target time $T$ does indeed affect the extent of amplification in perturbation energy observed, the dynamics at play qualitatively remains the same. The results presented here were obtained by fixing $T = 30$.

Although a majority of the studies on non-modal stability pertains to finding the optimal initial condition that extremises an objective functional (often the perturbation energy), such an undertaking for the current study is superfluous. By employing the direct-adjoint looping procedure, we are content with finding non-modal perturbations that are not necessarily global optimal initial conditions for these systems. 

\subsection{Results for two-layer flow}
In the previous section, it has been established two-layer shear flow is modally stable regardless of the value of the viscosity ratio $m$. Therefore, without loss of generality, we discuss results only for $m = 2$. It then follows that the lower layer is less viscous one. It has been verified that the principal features of the non-modal perturbations do not change for other values of $m$. 

\begin{figure}
  \centering
  \includegraphics[width=\textwidth]{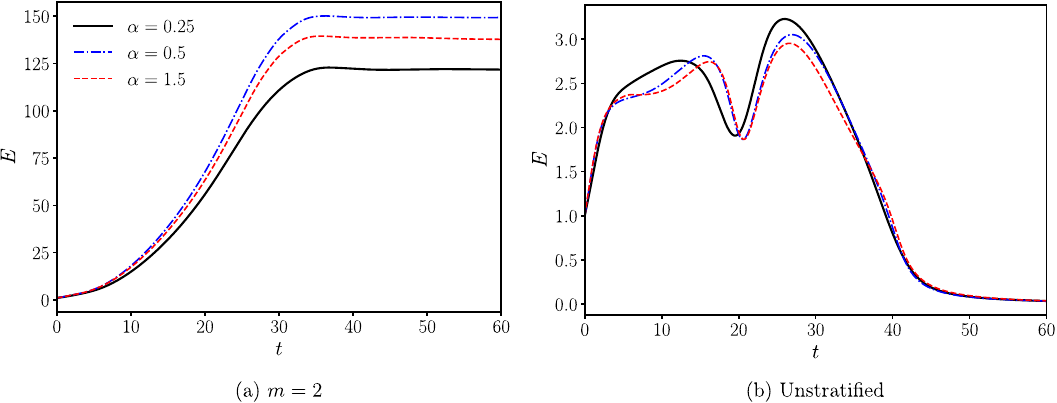}
  \caption{Perturbation energy $E$ as a function of time. (a) $m = 2$: for each value of wavenumber $\alpha$, the initial condition is evaluated by the direct-adjoint looping method. (b) The initial conditions used in (a) are evolved in the absence of viscosity stratification.}
  \label{fig:2L_E_v_t_alpran}
\end{figure}

The most striking feature of non-modal perturbations in shear flows is the transient algebraic amplification of perturbation energy $E$.  
In figure \ref{fig:2L_E_v_t_alpran} (a), the evolution of $E$ in time is shown for different wavenumbers.
After a period of energy amplification, $E$ is seen to become nearly constant at later times. In fact, there is an imperceptibly weak decay in $E$ that becomes more apparent at much later times. The extent of amplification seen here is significantly larger than in unstratified plane shear flow. This can be verified in two complementary ways. First, upon specifying the different layers to have equal viscosities while keeping all other conditions fixed and performing the optimisation procedure, we report that the maximum amplification seen is at most $\mathcal{O}(10)$. Next, we check what happens when the non-modal perturbation found for the stratified flow is introduced in a unstratified shear flow.
The resulting growth in $E$ is seen to be extremely weak and the perturbation dies out at later times (see figure \ref{fig:2L_E_v_t_alpran} (b)). 

\begin{figure}
  \centering
  \includegraphics[width=\textwidth]{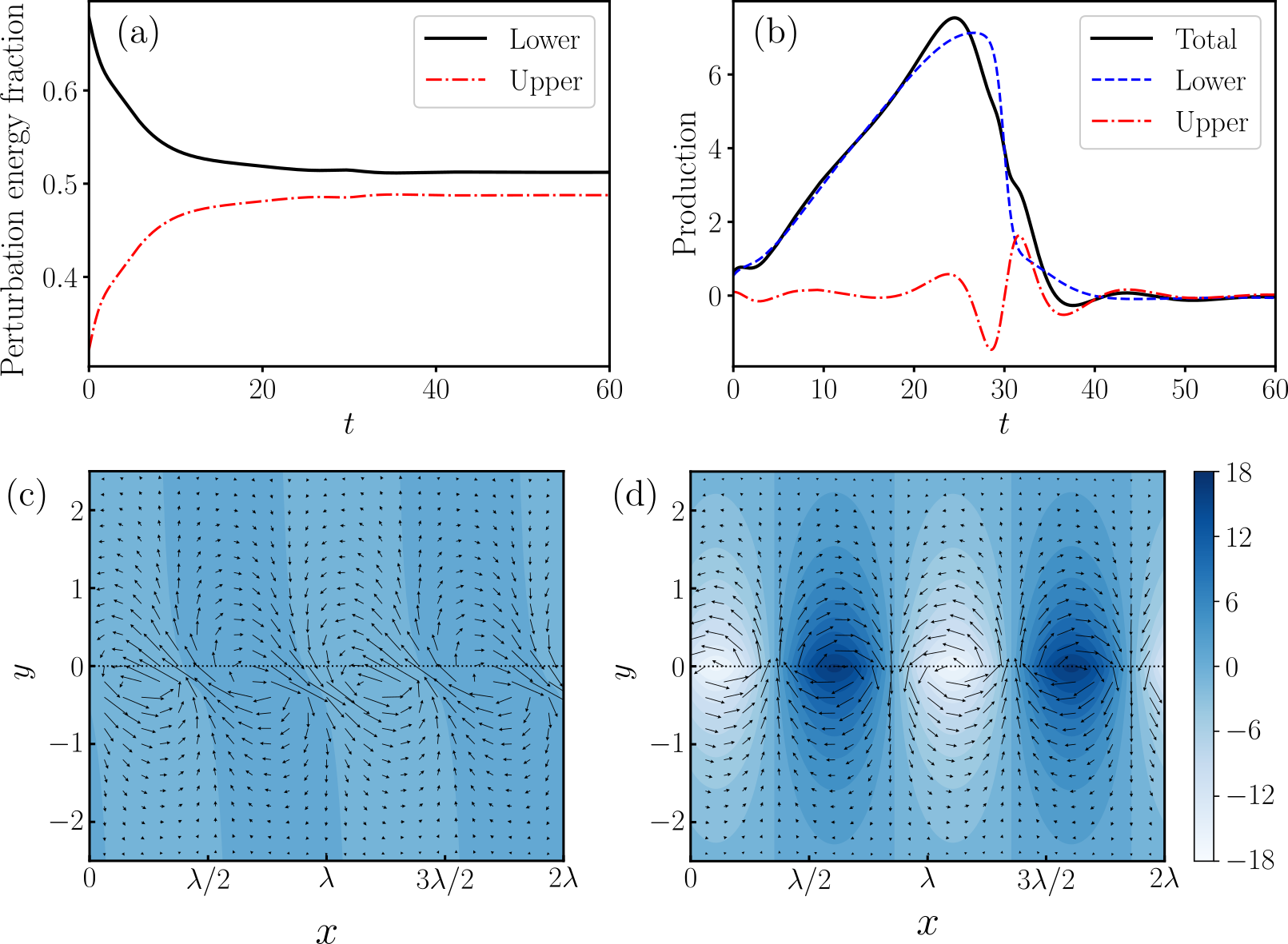}
  \caption{$m = 2$ and $\alpha = 1$. (a) Perturbation energy fraction and (b) production as a function of time. The perturbation structure at (c) $t = 0$ and (d) $t = 50$; the colour, the arrows and the dotted lines denote the streamfunction, the velocity vector and the interface respectively. For the arrows: (c) $l = 0.1638L$, (d) $l = 0.9968L$.}
  \label{fig:2L_nmod_coll}
\end{figure}

Now that we have seen the non-modal perturbations can exhibit considerable amount of transient growth, let us examine a specific case in greater detail.
For this purpose, we choose the non-modal perturbation for the case when $m = 2$ and $\alpha = 1$; we have verified that other choices for these parameters do not yield qualitatively different results. The distribution of perturbation energy is seen to be uneven between the two layers throughout the evolution (see figure \ref{fig:2L_nmod_coll} (a)). The disparity is greatest at $t = 0$, and reduces as time increases. As was observed for $E$ at later times, the perturbation energy content in the two layers too approach nearly constant values. 
The perturbation energy content in the lower (less viscous) layer is seen to be greater for the the period of the evolution considered here.

When it comes to kinetic energy production, it is seen that the differences between the two layers can be far more dramatic (see figure \ref{fig:2L_nmod_coll} (b)). For the phase of evolution where the perturbation energy is continuously increasing, it is seen that the production in the lower layer is significantly larger. The production in the upper (more viscous) layer is negative for certain periods. As the perturbation energy approaches the near saturation phase, the production in the lower layer starts becoming much lesser. At later stages, the difference in the production between the two layers is no longer as pronounced.

We next examine the structure of the perturbation at different times during its evolution.
Panels (c) and (d) in figure \ref{fig:2L_nmod_coll} show the perturbation structures at $t = 0$ and $t = 50$ respectively; note that the snapshot at $t = 50$ is representative of the perturbation at times after the instant the maximum in $E$ has occurred.
The initial perturbation is seen to be more pronounced in the less viscous lower layer. The streamfunction wavefronts are also initially tilted against the mean shear in both layers that allows for the Orr mechanism to occur to some extent. The perturbation field evolves such that the tilt of the streamfunction against the mean shear is no longer as considerable as it was at $t = 0$. The perturbation is found to be propagating downstream while its structure evolves in time. Near the interface, vortices can be discerned within the perturbation structure at later times. The peaks of the perturbation voriticty magnitude is observed to lie just below the interface in the lower (less viscous) layer. 

\subsection{Results for three-layer flow}
We now consider the three-layer plane shear flow that is modally unstable only when the middle layer is the least viscous. The additional layer gives us a larger parameter space to explore. We shall first discuss characteristics of non-modal perturbations in the core-annular set-up with the upper and lower layers having identical viscosities. We then proceed to examine mean flow configurations where all three layers have unequal viscosities; note that the viscosity of the middle layer no longer needs to be the largest in these flow configurations as we already established their modal stability in subsection \ref{ssec:3L_mod}. 

\begin{figure}
  \centering
  \includegraphics[width=0.5\textwidth]{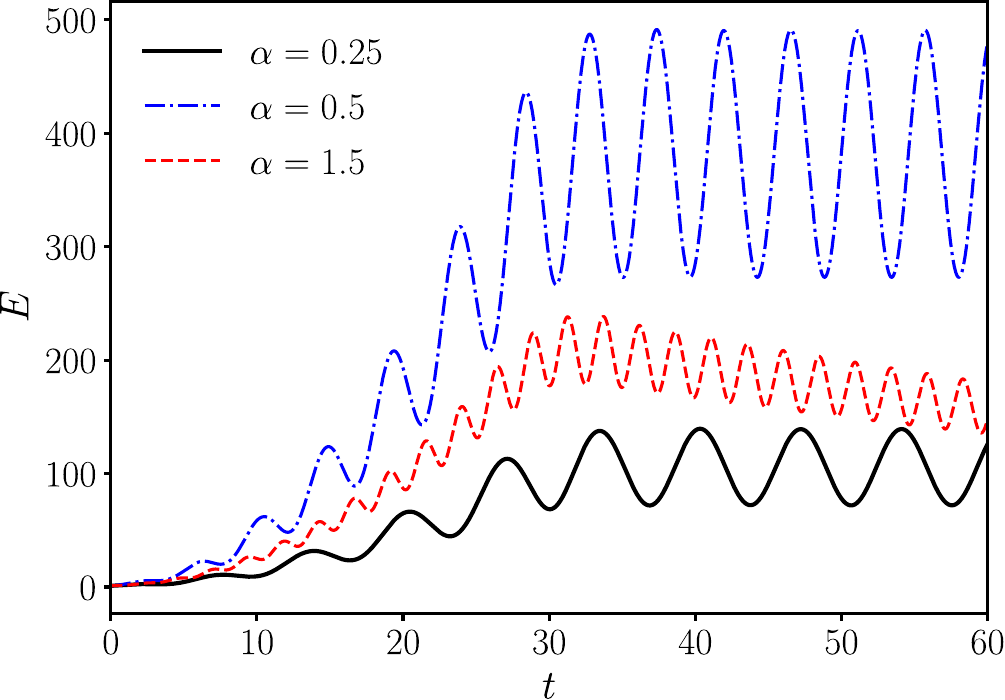}
  \caption{Evolution of perturbation energy for different wavenumbers. Base flow parameters: $m_l = m_u = 0.5$.}
  \label{fig:3L_E_v_t}
\end{figure}

\subsubsection{Core-annular configurations}
For the mean flow where the outer layers have equal viscosity, we set $m_l = m_u = 0.5$; there were no qualitative differences for other modally stable core-annular configurations. Figure \ref{fig:3L_E_v_t} shows how the energy $E$ evolves in time for different non-modal perturbations. The approach to the maximum value of $E$ is not monotonic with the curves revealing an oscillatory in time behaviour. The time periods between successive crests/troughs of $E$ settle close to a fixed value only at the later stages of the perturbation evolution. The number of oscillations in a given time window increases with the wavenumber of the perturbation.  In the figure, it might appear that $E$ decays at later time only for $\alpha = 1.5$ at first. We report that the perturbations eventually does decay for all values of $\alpha$ at later times. For the present, the focus will be on the early stage dynamics of the perturbations. For the sake of brevity, we refer to the phase of non-modal growth before $E$ attains its maximum value as the oscillatory growth phase. 

\begin{figure}
  \centering
  \includegraphics[width=\textwidth]{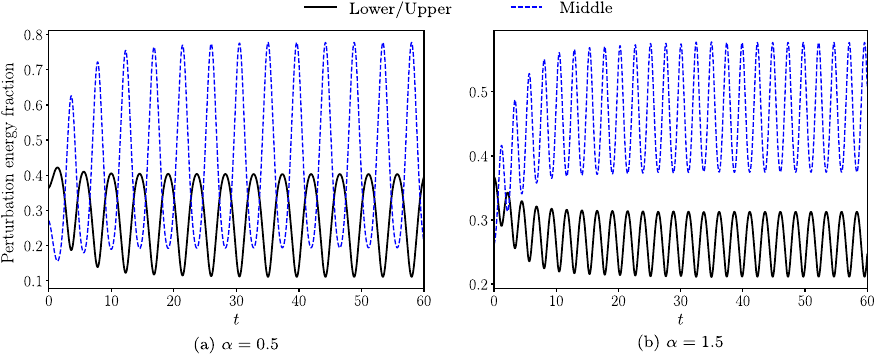}
  \caption{The evolution of perturbation energy distribution in the different layers. Base flow parameters: $m_l = m_u = 0.5$.}
  \label{fig:3L_Ef_v_t}
\end{figure}

For the above set of initial conditions, figure \ref{fig:3L_Ef_v_t} shows the distribution of perturbation energy among the different layers as a function of time. The oscillatory nature of the perturbation evolution is again reflected here.
In terms of perturbation energy content, the upper and lower layers are equivalent at all times. 
This should not come as a surprise owing to the symmetric nature of the mean flow. At $t = 0$, the initial condition can be seen to be more energetic in the less viscous outer layers. As the perturbation evolves in time, there are periods when the contribution of the perturbation energy in the middle layer dominates that of the outer layers put together. This behaviour is seen despite the middle layer having the smallest volume fraction and being the most viscous of the three layers.
Additionally, we note that the perturbation energy fraction of the middle layer (and subsequently, the outer layers) go through a wider range of values for lower values of the wavenumber over the course of the evolution. 

\begin{figure}
  \centering
  \includegraphics[width=\textwidth]{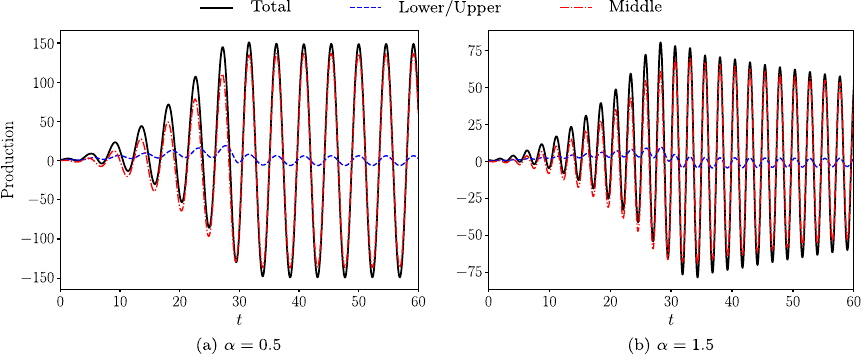}
  \caption{Production in the different layers as a function of time. Base flow parameters: $m_l = m_u = 0.5$.}
  \label{fig:3L_Prod_v_t}
\end{figure}

The picture becomes more interesting when we examine the production in the different layers (see figure \ref{fig:3L_Prod_v_t}). At first, it might seem appear that the evolution of the perturbation energy is solely tied to the dynamics in the middle layer. The contribution of the outer layers to the production is seen to be comparatively much smaller after all. 
During the oscillatory growth phase, it can be noted that the curves representing the total production always lie above those of the production in the middle layer. This implies that the production in the outer layers during this phase is always positive. In particular, when the production in the middle layer is negative, the production in the outer layers ensures that the perturbation energy does not go back to its original value after each oscillation. This slight offset appears to be crucial in allowing for the perturbation energy to grow, albeit in an oscillatory manner, during the initial stages. At later times, the production in the middle and outer layers end up being in phase for all wavenumbers. When $\alpha = 1.5$, the oscillatory decay of the production is in line with what was observed in the perturbation energy. 

\begin{figure}
  \centering
  \includegraphics[width=\textwidth]{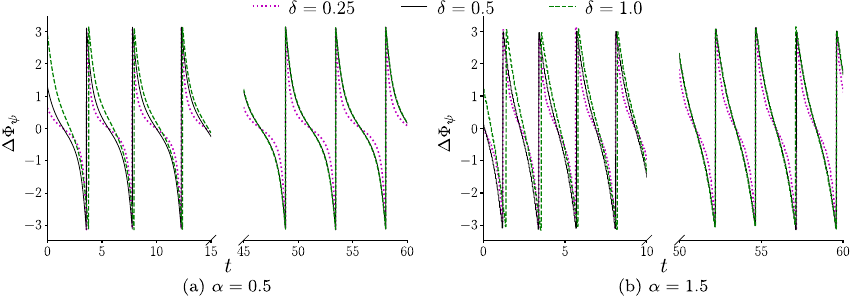}
  \caption{Phase difference of the streamfunction $\hat{\psi}$ at $y = \pm \delta$ as a function of time. Early ($t \in [0,15]$) and late ($t \in [45,60]$) stages of the evolution are shown. Base flow parameters: $m_l = m_u = 0.5$.}
  \label{fig:3L_phd_v_t}
\end{figure}

The discussion so far has been on volume averaged quantities. To take a closer look at the perturbation structure, we first examine the phase of the streamfunction $\Phi_{\psi}$ as a function of time and the vertical coordinate. In particular, to get a sense of how $\Phi_{\psi}$ varies about the mid-point of the central layer, we evaluate $\Delta \Phi_{\psi}(\delta,t) = \Phi_{\psi}(\delta,t) - \Phi_{\psi}(-\delta,t)$ for different values of $\delta$. Such a definition for $\Delta \Phi_{\psi}$ can be deemed natural owing to the symmetry of the mean flow. In figure \ref{fig:3L_phd_v_t}, the values of $\delta$ are selected such that different regions in the flow domain are represented; while the  $\delta = 0.5$ is representative of the interfaces, the curves for $\delta = 0.25$ and $\delta = 1.0$ are typical of the  middle and outer layers respectively. $\Delta \Phi_{\psi}$ are seen to be different during the oscillatory growth phase of the perturbation evolution. At later stages, $\Delta \Phi_{\psi}$ for $\delta = 0.5,1 .0$ become equal. The implication of this observation will be made clear and discussed in the subsequent. 

We now focus on the evolution of the non-modal perturbation for one specific wavenumber: $\alpha = 0.5$.
At $t = 0$, the perturbation streamfunction is inclined against the mean shear with the degree of inclination varying with $y$. This is favourable for the Orr mechanism.
Figures \ref{fig:3L_Prod_v_t} and \ref{fig:3L_phd_v_t} hinted at the importance of the dynamics in the outer layers during the initial stages of the perturbation evolution. Therefore, we start with observations made in the outer layers. Away from the interface, the gradient of $\Phi_{\psi}$ quickly reduces to zero in time.
The streamfunction fronts in the outer layer become nearly vertical in each of the outer layers. 
There is however a region in the immediate vicinity of the interfaces where $\Phi_{\psi}$ changes rapidly. This region becomes narrower quickly after the initial growth phase. This behaviour is reflected in the plots of $\Delta \Phi_{\psi}$ in figure \ref{fig:3L_phd_v_t}.

We next turn our attention to the dynamic processes in the middle layer, which is shown to have considerable dominance in the production of perturbation energy (see figure \ref{fig:3L_Prod_v_t}). The Orr mechanism certainly comes into play. However, the novel distinction in the current setting is that once the energy approaches a local minima in time, the streamfunction wavefront is seen to be tilted against the mean shear yet again ensuring another round of Orr mechanism dynamics. 
At early times, $E$ at the end of each Orr mechanism cycle is always greater than what it was the start. Recall the important role of the production in the outer layers in the oscillatory growth stage (see figure \ref{fig:3L_Prod_v_t}).
For the core-annular mean flow configurations considered here, this phenomenon is seen to occur regardless of the perturbation wavenumber.

\begin{figure}
  \centering
  \includegraphics[width=\textwidth]{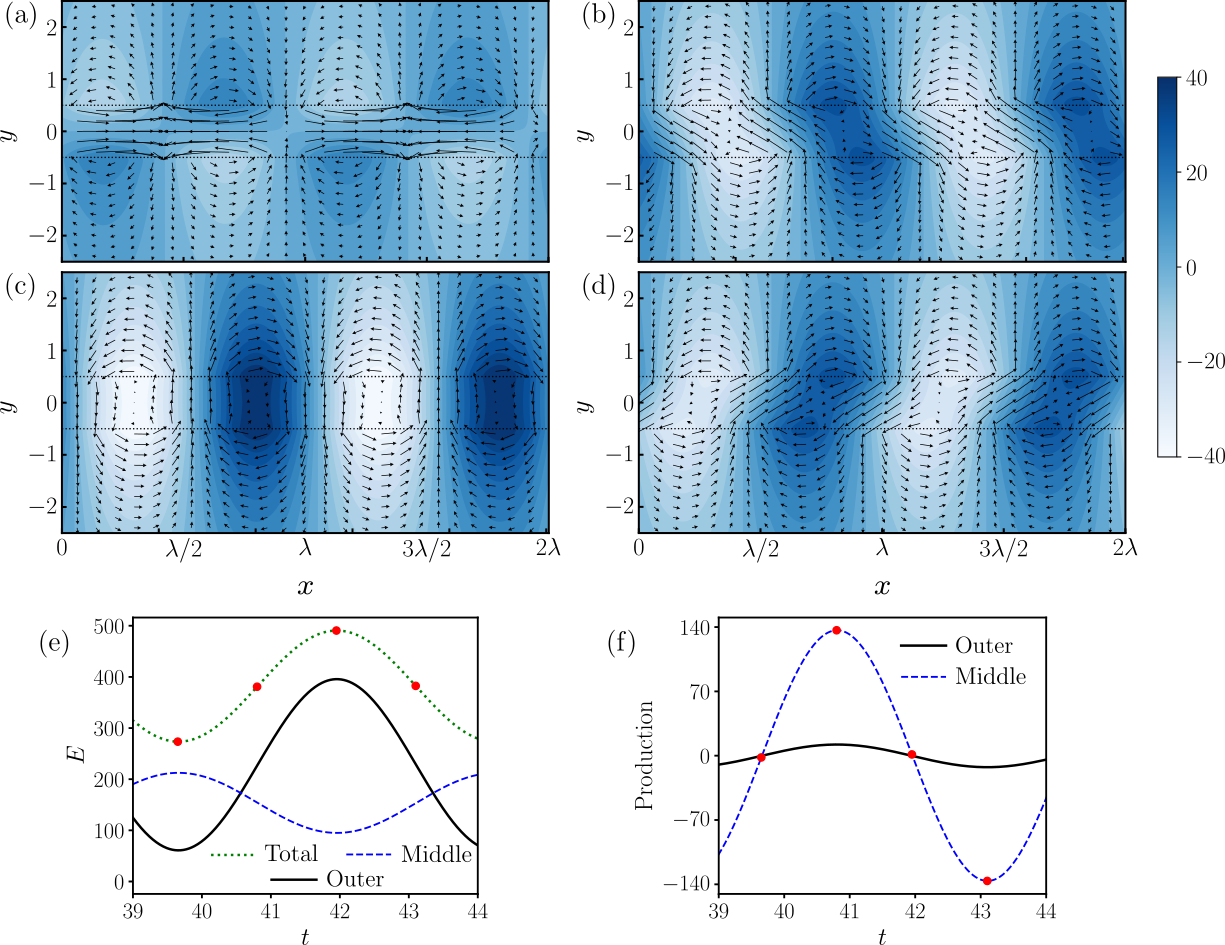}
  \caption{When $m_l = m_u = 0.5$ and $\alpha = 0.5$, the perturbation structure is shown at (a) $t = 39.65$, (b) $t = 40.80$, (c) $t = 41.95$, (d) $t = 43.10$. The colour, the arrows and the dotted lines denote the streamfunction, the velocity vector and the interfaces respectively; for the arrows: (a) $l = 0.9999L$, (b) $l = 0.8470L$, (c) $l = 0.6613L$, (d) $l = 0.8401L$. In (e) and (f), the red dots correspond to the snapshots shown in (a)-(d).}
  \label{fig:3L_pert_coll_1}
\end{figure}

The perturbation structures at four different times during the later stages of its evolution are shown in panels (a)-(d) of figure \ref{fig:3L_pert_coll_1}.
Panels (a) and (c) correspond to a local minima and maxima of the perturbation energy in time respectively (see figure \ref{fig:3L_pert_coll_1} (e)). When $E$ is at a minima, what immediately catches the eye are alternating sets of two streamwise jets converging in the middle layer. It is notable that the largest perturbation velocity magnitudes correspond to the minimum of $E$. On the other hand, when $E$ is at a maxima, the perturbation resembles a roll centred about the middle layer. At this point, the perturbation is seen to more energetic in the outer layers. The panels (b) and (d) correspond to times when the production is maximum and minimum (see figure \ref{fig:3L_pert_coll_1} (f)). As was established earlier, the middle layer alone accounts for most of the production. In this regard, when we focus only on the perturbation structure in the middle layer, the orientation of the streamfunction wavefronts in panels (b) and (d) are typical of those seen in the classical Orr mechanism during the early and late stages respectively. With increase in time, these perturbation structures emerge again cyclically. 

We report that the broad features of the structures shown in figure \ref{fig:3L_pert_coll_1} are seen repeatedly throughout the perturbation evolution in the core-annular flow configuration. This can be discerned even during the early stages of the perturbation evolution. The prevalence of these structures allows for a novel regenerative form of the Orr mechanism. During the early stages, this regenerative Orr mechanism allows for the perturbation energy to increase significantly. At later stages, it allows for the perturbation to remain in a state of higher energy for periods far longer than what would be observed in uniform plane shear flows. 

\subsubsection{Layers with unequal viscosities}
The most general configurations of the mean flow have the three fluid layers being of unequal viscosities. We consider two specific mean flows here: (I) $m_l = 0.6$ and $m_u = 0.8$, (II) $m_l = 2$ and $m_u = 0.5$. In configuration I, the middle layer is the most viscous of the three. On the other hand, the upper layer takes the mantle of being the most viscous in configuration II. 

\begin{figure}
  \centering
  \includegraphics[width=\textwidth]{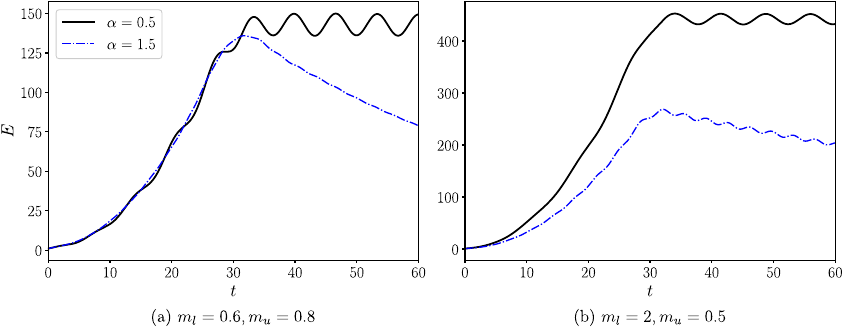}
  \caption{Evolution of perturbation energy for different wavenumbers. The upper and lower layers have unequal viscosities.}
  \label{fig:3L_E_v_t_uneq_conf}
\end{figure}

Figure \ref{fig:3L_E_v_t_uneq_conf} shows how the perturbation energy changes as a function of time for two different wavenumbers. Once again, the non-modal perturbations demonstrate significant amplification in $E$. Unlike what was observed for core-annular configurations, $E$ initially increases monotonically in time. The late-time behaviour is dependent on the wavenumber of the perturbation. For the two mean flows considered, the perturbations with $\alpha = 0.5$ have a nearly constant oscillatory evolution stage after the initial transient. For larger wavenumbers ($\alpha = 1.5$), as was seen in the core-annular configurations, the curves of $E$ show an oscillatory decay after attaining maximum transient growth. The curves in figure \ref{fig:3L_E_v_t_uneq_conf} share some of the features already reported for the core-annular three-layer flow and the two-layer flow. We will elaborate on this aspect in the following by restricting our discussion to perturbations with $\alpha = 0.5$.

\begin{figure}
  \centering
  \includegraphics[width=\textwidth]{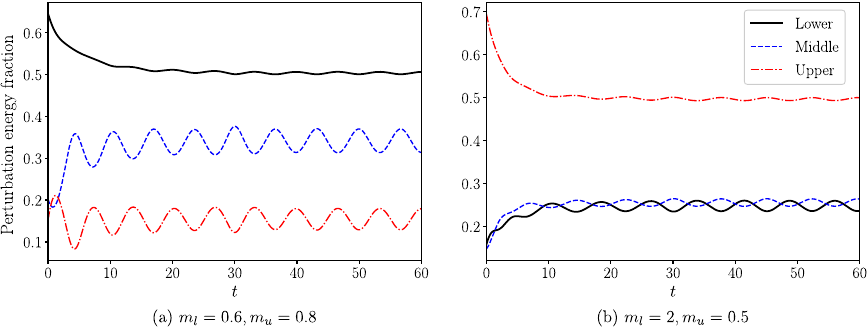}
  \caption{Perturbation energy distribution in the different layers as a function of time when $\alpha = 0.5$. The lower and upper layers are the least viscous in (a) and (b) respectively.}
  \label{fig:3L_Ef_v_t_uneq_conf}
\end{figure}

The non-modal perturbation does not have any form of symmetry along the vertical coordinate owing to the nature of the mean flows considered. This is immediately apparent when we plot the perturbation energy fraction as a function of time in figure \ref{fig:3L_Ef_v_t_uneq_conf}. The least viscous layer is the largest contributor to $E$ for all time, a characteristic also seen in the non-modal perturbations obtained for the two-layer flow (see figure \ref{fig:2L_nmod_coll} (a)). The perturbation is seen to be considerably strong in the middle layer, regardless of whether it is the most viscous or not. In particular, this can be seen in figure \ref{fig:3L_Ef_v_t_uneq_conf} (a). The undulations seen in the curves resemble those seen for non-modal perturbations in the core-annular setting. However, it is pointed out that the undulations are not nearly as prominent as was seen in figure \ref{fig:3L_Ef_v_t}.

\begin{figure}
  \centering
  \includegraphics[width=\textwidth]{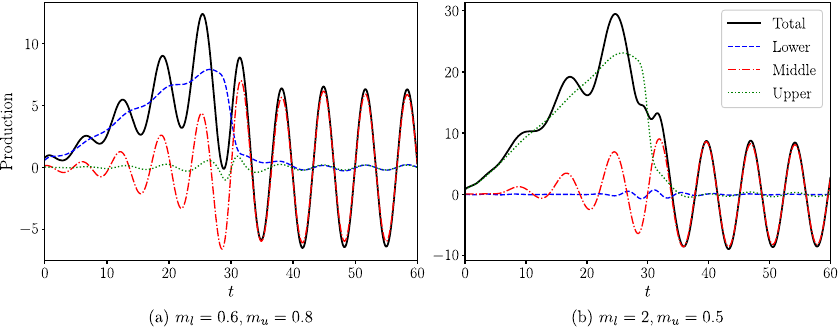}
  \caption{Production in the different layers as a function of time when $\alpha = 0.5$. The lower and upper layers are the least viscous in (a) and (b) respectively.}
  \label{fig:3L_Prod_v_t_uneq_conf}
\end{figure}

Figure \ref{fig:3L_Prod_v_t_uneq_conf} shows the evolution of the production and its subdivision between the different layers.
The least viscous layer drives the production during the early stages of the perturbation evolution (till $t \approx 30$).
In this regard, there is a marked difference from what was observed in the core-annular setting (see figure \ref{fig:3L_Prod_v_t}). On the other hand, the early-time dynamics is similar to what was seen in the two-layer flow (see figure \ref{fig:2L_nmod_coll} (b)). The production in the middle layer slowly picks up before eventually dominating at later times; the production in the outer layers are negligible at this stage. At later times, comparisons with the core-annular configurations are once again more suitable. Nonetheless, it is important to recall that the least viscous layer continues to account for the largest portion of the perturbation energy content.

\begin{figure}
  \centering
  \includegraphics[width=\textwidth]{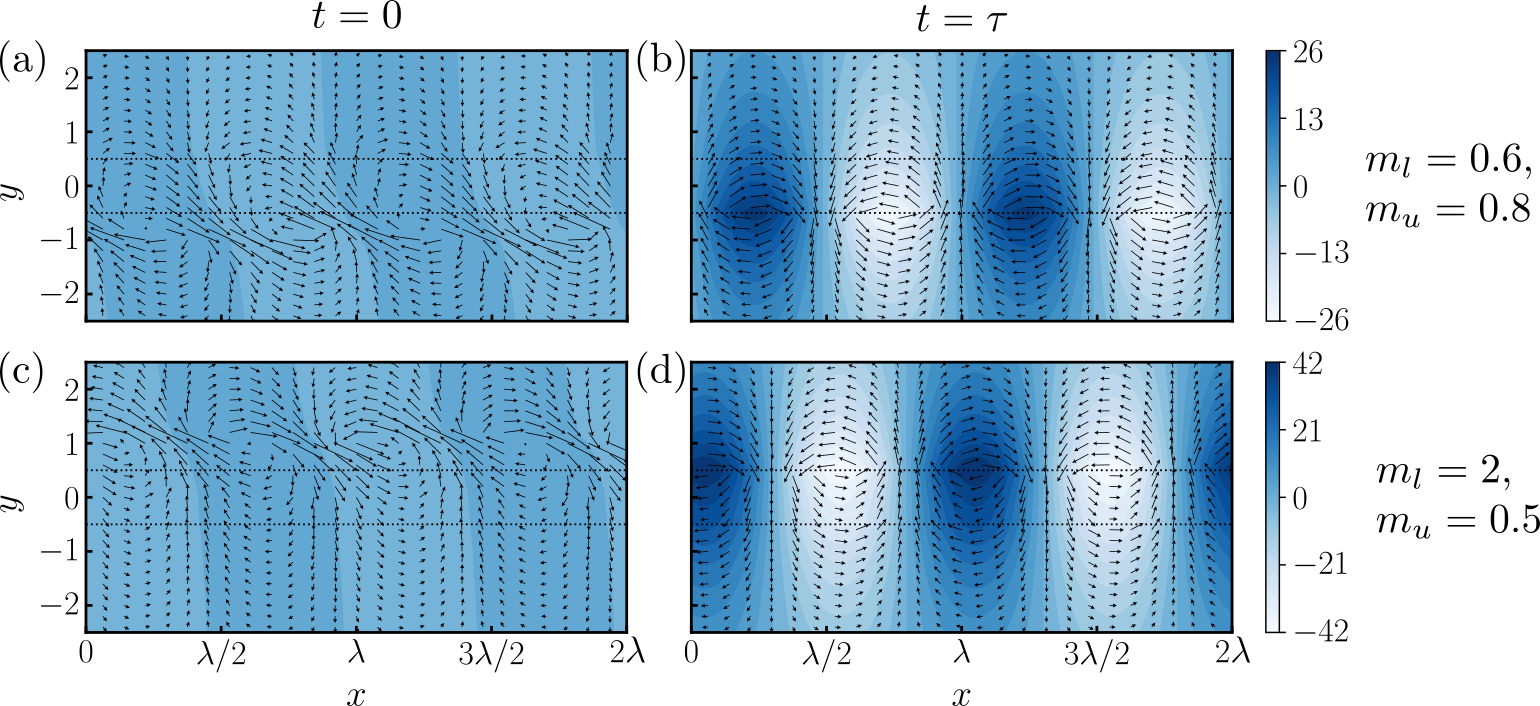}
  \caption{The perturbation structure at $t = 0$ and $t = \tau$ for different mean flows. The colour, the arrows and the dotted lines denote the streamfunction, the velocity vector and the interfaces respectively. Top row: $\tau = 46.65$; arrow length scale: (a) $l = 0.1562L$, (b) $l = 0.9171L$. Bottom row: $\tau = 41.40$; arrow length scale: (c) $l = 0.1019L$, (d) $l = 0.9996L$. $L$ is different in the top and bottom rows.}
  \label{fig:3L_uneq_conf_coll}
\end{figure}

We now proceed to discuss the structure of the non-modal perturbations.
In figure \ref{fig:3L_uneq_conf_coll}, we show snapshots at $t = 0$ and at an instant $t = \tau$ when $E$ is maximum. It is to be noted that $\tau$ need not be uniquely defined for any given non-modal perturbation (see figure \ref{fig:3L_E_v_t_uneq_conf}).
The similarity of these perturbations to the ones seen in the two-layer flow is more readily seen now. The perturbation is more stronger in the least viscous layer with the initial tilt of the streamfunction fronts allowing for the Orr mechanism to amplify the perturbation energy at early times. 
While discussing figures \ref{fig:3L_E_v_t_uneq_conf} and \ref{fig:3L_Prod_v_t_uneq_conf}, it was suggested earlier that the dynamics at later times bears some similarity to the regenerative Orr mechanism seen in the core-annular configurations. However, upon examining different snapshots visually, it is not straightforward to discern the process in terms of the tilt of the streamfunction wavefronts against the mean shear. 

\subsubsection{Perturbation evolution in unstratified flow}
\begin{figure}
  \centering
  \includegraphics[width=0.45\textwidth]{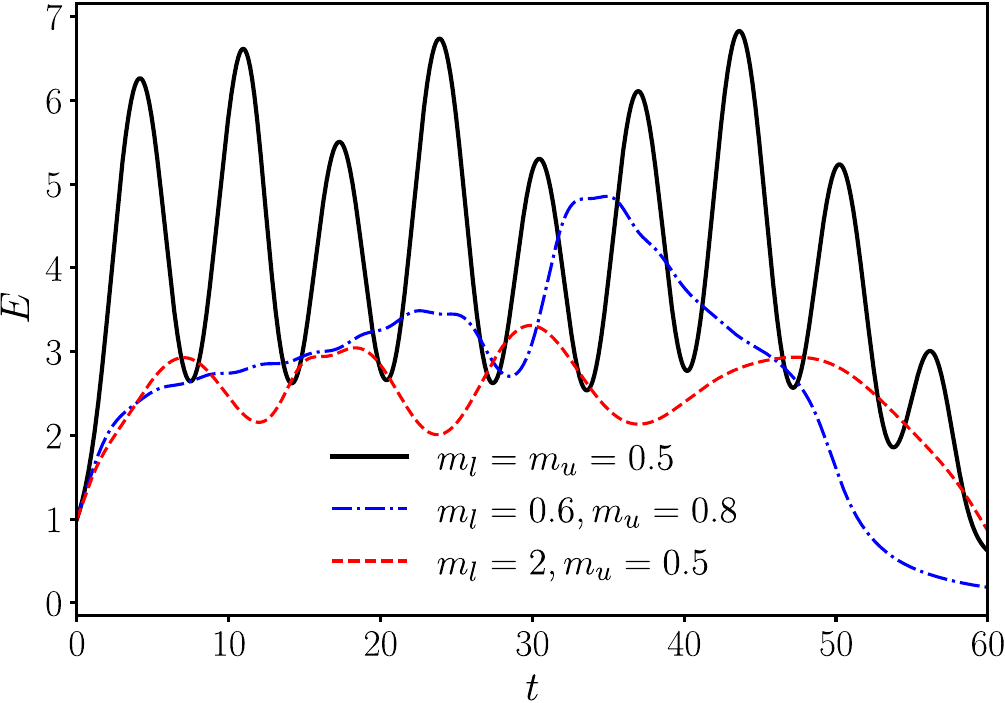}
  \caption{Evolution of perturbation energy when the mean flow is unstratified. The perturbation wavenumber is $\alpha = 0.5$, and the initial conditions obtained for the stratified flow are used here.}
  \label{fig:3L_Unstr_E_vs_t}
\end{figure}

Finally, we will briefly address how the non-modal perturbations found for the stratified three-layer plane shear flow would evolve when introduced in an unstratified plane shear flow.
For the sake of brevity, we only report the results for those specific non-modal perturbations which were analysed in greater depth earlier in this section. We emphasise that similar trends are observed for other modally stable mean flow configurations and perturbation wavenumbers. 
Figure \ref{fig:3L_Unstr_E_vs_t} shows the evolution of the perturbation energy when $\alpha = 0.5$. 
Once again, as previously seen in figure \ref{fig:2L_nmod_coll} (b), it is abundantly clear that the mean shear being non-uniform is vital for observing the levels of perturbation energy amplification reported above. 

\subsection{Connection to discrete modal eigenfunctions}
Can we state something more precise with regard to the initial conditions obtained from the optimisation procedure? It turns out that we can when we turn to insight gained from modal analysis of these systems. The base flows under consideration are smoothened counterparts of sharp interface configurations for which they would have been discontinuous jumps of the mean shear rate. For such piecewise linear profiles, the number of elements in the discrete spectrum is given by the number of interfaces. When working with these smoothened mean flow profiles, these discrete modes are no longer present. However, their effects can still manifest by means of a set of continuous spectrum modes in the form of a quasi-mode \citep{Briggs_etal_1970PF,Schecter_etal_2000PF,Shrira_Sazonov_2001JFM,Dixit_Govindarajan_2011JFM}.

We shall continue the discussion using the non-modal perturbations highlighted in the preceding subsections. For the discrete modes of the piecewise continuous flow, we can identify the critical layer where the piecewise continuous mean flow velocity matches the phase speed of each discrete mode. For piecewise continuous flows, the critical layer is not as vital as $U'' = 0$ everywhere except at the interfaces. We report that there is very little difference in the critical layer locations regardless of whether the flow is bounded or unbounded in the current setting. If we were to use the smoothened mean flow instead, the evaluated location again barely changes. For the configurations studied here, the phase speeds and the corresponding critical layer locations for the bounded piecewise linear mean flow are given in table \ref{tab:crit_lay}.

\begin{table}[b]
  \caption{\label{tab:crit_lay}
    Phase speeds and locations of critical layer for discrete modal perturbations of the piecewise continuous mean flow. 2L and 3L refer to two-layer and three-layer flows respectively. For 3L systems, only the critical layer in the less viscous layer is given.
  }
\begin{ruledtabular}
\begin{tabular}{lccc}
\textrm{Mean flow parameters} &
$\alpha$ & $c$ & $y_c$ \\
\colrule
2L: $m = 2$ & 1 & -0.333322 & -0.249992\\
3L: $m_l = m_u = 0.5$ & 0.5 & $\pm 1.361278$ & $\pm 0.930639$\\
3L: $m_l = 0.6$, $m_u = 0.8$ & 0.5 & -1.125263 & -0.875158\\
3L: $m_l = 2$, $m_u = 0.5$ & 0.5 & 1.594689 & 1.047344\\
\end{tabular}
\end{ruledtabular}
\end{table}

\begin{figure}
  \centering
  \includegraphics[width=\textwidth]{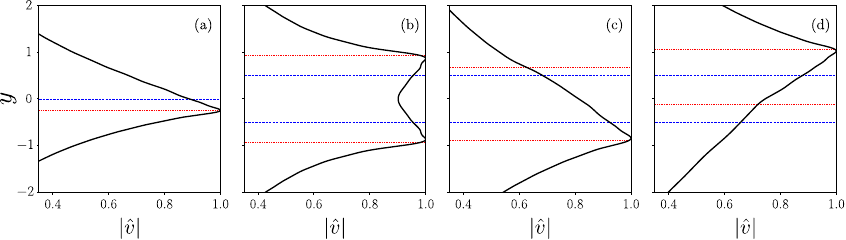}
  \caption{Profiles of $\left|\hat{v}\right|$ when $\alpha = 0.5$ for different flow configurations: (a) two-layer flow with $m = 2$, and three-layer flows with (b) $m_l = m_u = 0.5$, (c) $m_l = 0.6$, $m_u = 0.8$, (d) $m_l = 2$, $m_u = 0.5$. The blue-dashed lines represent the interfaces. The red-dotted lines identify the critical layers given in table \ref{tab:crit_lay}.}
  \label{fig:QM_analy}
\end{figure}

Upon examining the initial conditions more closely, we find the critical layers to be located in close proximity to the peak of different perturbation quantities. As an illustration, we plot the magnitude of the normal component of the perturbation velocity ($\left|\hat{v}\right|$) in figure \ref{fig:QM_analy}. For three-layer flows, the critical layer that determines the localisation of the perturbation is the one that lies in the layer with the largest shear rate, i.e., the least viscous layer. We also report that the region around the critical region continues to be energetically active at later times albeit not in the dominant fashion seen at $t = 0$.

We emphasise that the optimisation procedure employed here does not specify the perturbation to be more stronger in the vicinity of the critical layers or at any specific location. Such initial conditions emerge after the first couple of iterations of the algorithm. On the other hand, one might be able to construct initial conditions centred about the critical layer in an ad hoc manner. But such prescriptions can not ensure the levels of perturbation amplification seen in the results here. For example, \citet{Polyachenko_Shukhman_2022PF} show that an initial vorticity perturbation in the form of a Gaussian first undergoes Landau damping followed by an asymptotic algebraic decay without ever entering a phase of transient growth. 

\section{Conclusion and outlook}
\label{sec:conc}
The inviscid stability characteristics of plane shear flows comprising two and three layers to 2D perturbations are examined in this study. Every configuration of the two-layer flow is shown to be modally stable. On the other hand, for the three-layer flow, modal instabilities emerge when the middle layer has the lowest viscosity. Upon analysing the piecewise continuous counterpart of these shear flows, we are able to show that a minimum of three layers in the mean flow is a prerequisite for modal instability. The corresponding perturbation structure highlight the importance of the interfacial regions.

Most of the focus of this work is on the non-modal stability characteristics of these flows. We report significant differences from what is observed for the shear flow of a single fluid layer. The amplification of perturbation energy is considerably larger in both shear flows of two and three layers for a wide range of wavenumbers. Moreover, these non-modal perturbations do not decay quickly. Moreover, the evolution of the perturbation energy is not necessarily monotonic in the growth phase for the three-layer flow. The same initial conditions when introduced in an unstratified shear flow certainly do not evolve to replicate the levels of the transient growth seen in the layered flows. We posit these observations to be among the more important take-away messages from this work.

The structure and the subsequent dynamics of the non-modal perturbation is dependent on the mean flow considered. For the two-layer flow, the perturbation energy is shown to be always larger in the less viscous layer. The less viscous layer is also the more dominant contributor to the production during the phase of perturbation energy growth. At later stages, the net production almost reduces to zero resulting in the apparent plateauing of the perturbation energy.

In core-annular three-layer flow configurations, the optimised non-modal perturbations are energetically equivalent in the outer layers. The corresponding dynamics result in the middle layer, which has the smallest volume fraction and is the most viscous, accounting for more perturbation energy for a considerable portion of time. At first glance, the middle layer appears to dominate the production of perturbation energy throughout the perturbation evolution. However, the outer layers end up subtly playing a vital role during the oscillatory growth phase. Throughout the perturbation evolution, a novel regenerative form of the Orr mechanism is in evidence.

When the three layers are distinct from each other, the non-modal perturbations demonstrate some features of those obtained for both the two-layer and the core-annular three-layer flows. As in the two-layer flow, the least viscous layer accounts for most of the perturbation energy. During the early stages of the evolution, the least viscous layer also dominates the production of perturbation energy. At the same time, the participation of the middle layer in the production process picks up. Comparisons with the three-layer core-annular flow configurations are more apt at later stages of the perturbation evolution when the middle layer ends up dominating the production. However, the level of production is not to the same extent resulting in smaller oscillations of the perturbation energy in time.

This study sheds light on the dynamics of a class of non-modal perturbations in viscosity stratified shear driven flows. Such scenarios can be envisaged in large geophysical flows where a stratification of the eddy viscosity may be considered \citep{Constantin_Johnson_2019BLM,Dritschel_etal_2020OS}. The fact that the mean shear is not constant everywhere leads to scenarios where the perturbation energy amplification is considerably significant even for 2D perturbations. In this regard, it would be interesting to verify if the phenomena discussed here can play a more direct role in the nonlinear evolution of such flows. This might perhaps be in conjunction or in opposition to 3D processes, which are typically more potent in rendering the flow nonlinear in canonical shear flows.

The very nature of the mean flows considered here offer a large parameter space which remains to be explored. For instance, altering the depths of the different layers offer one such avenue for future study.
If density stratification effects were to be included, it is possible to get other forms of non-modal perturbations where the initial forcing is primarily due to buoyancy \citep{Jose_etal_2015PRSA,Jose_etal_2018FDR} bringing in an interplay between perturbation potential and kinetic energies \citep{Farrell_Ioannou_1993JAS,Kaminski_etal_2014JFMR}.
In an upcoming work, the consequences of including viscous and diffusive effects in the full nonlinear setting will be discussed. The results presented here will serve as a reference for studies that incorporate some of the factors mentioned above.

\appendix
\section{Computational method}
\label{app:num_meth_eval}
The computational domain is taken to be $y \in [-0.5\Sigma,0.5\Sigma]$. For this particular study, it is more desirable to have more grid points around $y = 0$. For this purpose, we shall make use of a stretched grid \citep{Vinokur_1983JCP,Raghav_etal_2022DAO}:
\begin{align}
  &y = \frac{a}{\sinh \left(bY_0\right)}\left\{\sinh \left[ b\left( \frac{Y}{\Sigma} - Y_0 \right)\right] + \sinh \left(bY_0\right)\right\} - \frac{\Sigma}{2}\label{eq:str_grd} \\
  \textrm{where }\;\; &Y_0 = \frac{1}{2b}\log \left[\frac{\Sigma + \left(\mathrm{e}^b - 1\right)a}{\Sigma + \left(\mathrm{e}^{-b} - 1\right)a}\right] \nonumber
\end{align}
In the above, we have introduced another coordinate $Y \in [0,\Sigma]$ that has a one-to-one mapping to the $y$ coordinate. Gauss--Lobatto collocation points are defined to discretise $Y$ as:
\begin{align}
  Y_j = \frac{\Sigma}{2}\left(1 + \cos \frac{\pi j}{N}\right),\;\; j = 0,~1,~\dots,~N \label{eq:GL_col_pts}
\end{align}
In equation \eqref{eq:str_grd}, $a$ is the location in the $\xi$ coordinate where the density of points is made higher; for all the results shown here, $a = 0.5\Sigma$. $b$ is a stretching parameter, whose value is selected such that at least 3 grid points lie in each of the mixing layers.

The discretised versions of the operators in equation \eqref{eq:Ray_mod} and \eqref{eq:Ray_ft} are obtained by employing the chain rule (using equation \eqref{eq:str_grd}) in conjunction the suitably scaled differentiation matrices for the collocation grid \citep{Weideman_Reddy_2000ACM}. Homogeneous boundary conditions are imposed on the normal velocity component at $y = \pm 0.5\Sigma$.  

For both the modal and non-modal analysis, the numerical results were obtained with $N = 401$ and $\Sigma = 11$. It is verified that the results do not change upon increasing the number of discretisation points. Likewise, the resulting changes in the results are not significant when $\Sigma$ is allowed to take on larger values.

\section{Dispersion relations of piecewise continuous linear flows}
\label{app:PL_mod}
Dispersion relations for piecewise continuous linear flows are derived by enforcing matching conditions of the perturbation pressure and the normal component of the perturbation velocity at the interfaces. In each layer of the piecewise continuous linear flow, the Rayleigh equation (\ref{eq:Ray_mod}) reduces to:
\begin{align}
  \left(\frac{\textrm{d}^2}{\textrm{d}y^2} - \alpha^2\right)\tilde{v} = 0. \label{eq:Ray_mod_PL}
\end{align}
Given the form of $\tilde{v}$, the perturbation pressure $\tilde{p}$ is given by:
\begin{align}
  \tilde{p} = -\frac{i}{\alpha}(U - c)\frac{\textrm{d}\tilde{v}}{\textrm{d}y} + \frac{i}{\alpha} U'\tilde{v}. \label{eq:pr_mod}
\end{align}
In each layer, we seek a solution $\tilde{v}$ satisfying conditions at the boundary/interface in addition to the equation \eqref{eq:Ray_mod_PL}. 

For brevity, only the dispersion relations for the bounded and unbounded flows are given as the procedure to obtain them is fairly standard \citep{Drazin_Reid_2004book,Schmid_Henningson_2001book}. Also note that the phase speed $c$ given below is with respect to a frame of reference where $U(0) = 0$.

\subsection{Two-layer flows}
For the bounded flow, where the depths of the lower and upper layers are given by $d_l$ and $d_u$ respectively, the dispersion relation is:
\begin{align}
  c = \frac{\left(1 - m\right)\left(1 - e^{-2\alpha d_l}\right)\left(1 - e^{-2\alpha d_u}\right)}{\alpha\left(1 + m\right)\left(1 - e^{-2\alpha \left(d_l + d_u\right)}\right)}. \label{eq:DR2L_bnd}
\end{align}
In the current work, $d_l = d_u = 5.5$. As $d_l,d_u \rightarrow \infty$, the dispersion relation reduces to:
\begin{align}
  c = \frac{\left(1 - m\right)}{\alpha\left(1 + m\right)}. \label{eq:DR2L_ubnd}
\end{align}
With $d_l = d_u = 5.5$, the values of $c$ obtained equations \eqref{eq:DR2L_bnd} and \eqref{eq:DR2L_ubnd} are extremely close. Only real values of $c$ exist, and therefore all are neutral modes. 

\subsection{Three-layer flows}
For the flow in the bounded setting, the dispersion relation becomes:
\begin{align}
  \mathcal{A}_bc^2 + \mathcal{B}_bc &+ \mathcal{C}_b = 0, \label{eq:DR3L_bnd} \\
  \textrm{with}\;  \mathcal{A}_b =~& 1, \nonumber \\
  \mathcal{B}_b =~& -\frac{\left[1 - e^{\alpha\left(1 - 2d_l\right)}\right]\left[1 - e^{-\alpha\left(1 + 2d_u\right)}\right]\Delta\mathcal{S}_{l} + \left[1 - e^{-\alpha\left(1 + 2d_l\right)}\right]\left[1 - e^{\alpha\left(1 - 2d_u\right)}\right]\Delta\mathcal{S}_{u}}{2\alpha\left[1 - e^{-2\alpha\left(d_l + d_u\right)}\right]}, \nonumber \\
  \textrm{ and }\mathcal{C}_b =~& -\frac{1}{4} + \frac{\left[1 - e^{\alpha\left(1 - 2d_l\right)}\right]\left[1 - e^{-\alpha\left(1 + 2d_u\right)}\right]\Delta\mathcal{S}_{l} - \left[1 - e^{-\alpha\left(1 + 2d_l\right)}\right]\left[1 - e^{\alpha\left(1 - 2d_u\right)}\right]\Delta\mathcal{S}_{u}}{4\alpha\left[1 - e^{-2\alpha\left(d_l + d_u\right)}\right]} \nonumber \\
  &- \frac{e^{-2\alpha}\Delta\mathcal{S}_l\Delta\mathcal{S}_u\left[1 - e^{\alpha\left(1 - 2d_l\right)}\right]^2\left[1 - e^{\alpha\left(1 - 2d_u\right)}\right]^2}{4\alpha^2\left[1 - e^{-2\alpha\left(d_l + d_u\right)}\right]^2} \nonumber \\
  &+ \frac{\Delta\mathcal{S}_l\Delta\mathcal{S}_u\left[1 - e^{\alpha\left(1 - 2d_l\right)}\right]\left[1 - e^{-\alpha\left(1 + 2d_l\right)}\right]\left[1 - e^{\alpha\left(1 - 2d_u\right)}\right]\left[1 - e^{-\alpha\left(1 + 2d_u\right)}\right]}{4\alpha^2\left[1 - e^{-2\alpha\left(d_l + d_u\right)}\right]^2}. \nonumber
\end{align}
In the above, $\Delta\mathcal{S}_{l}$ ($\equiv 1 - m_l^{-1}$) and $\Delta\mathcal{S}_{u}$ ($\equiv m_u^{-1} - 1$) are the jumps in the mean shear rate at the lower and upper interfaces respectively. As $d_l,d_u \rightarrow \infty$, the dispersion relation reduces to:
\begin{align}
  &\mathcal{A}c^2 + \mathcal{B}c + \mathcal{C} = 0, \label{eq:DR3L_ubnd} \\
  \textrm{with}\; &\mathcal{A} = 4\alpha^2,~\mathcal{B} = -2\alpha\left(\Delta\mathcal{S}_{l}+\Delta\mathcal{S}_{u}\right),\textrm{ and }\mathcal{C} = -\alpha^2 + \alpha\left(\Delta\mathcal{S}_{l} - \Delta\mathcal{S}_{u}\right) + \Delta\mathcal{S}_{l}\Delta\mathcal{S}_{u}\left(1 - e^{-2\alpha}\right). \nonumber
\end{align}
In the current study, where $d_l = d_u = 5$, the differences in the values of $c$ obtained equations \eqref{eq:DR3L_bnd} and \eqref{eq:DR3L_ubnd} are not significant. 

Some additional insight may be obtained directly by analysing the dispersion for the unbounded flow (equation \eqref{eq:DR3L_ubnd}). In this regard, it is useful to define:
\begin{align}
  m_l = 1 + \delta_l, m_u = 1 + \delta_u.
\end{align}
As $m_l, m_u > 0$, it follows that $-1 < \delta_l,\delta_u < \infty$. The discriminant $\mathcal{D}$ of equation \eqref{eq:DR3L_ubnd} is given by:
\begin{align}
  \mathcal{D}  = \frac{4\alpha^2(\delta_l - \delta_u)^2}{m_l^2m_u^2} + 16\alpha^4 - \frac{16 \alpha^3 (\delta_lm_u + \delta_u)}{m_lm_u} + \frac{16\alpha^2\delta_l\delta_u(1 - e^{-2\alpha})}{m_lm_u}
\end{align}
Exponentially unstable modes exists when $\mathcal{D} < 0$; also note that as $\alpha > 0$, $(1 - e^{-2\alpha}) > 0$. When $0 < m_l, m_u < 1$, or equivalently when $-1 < \delta_l, \delta_u < 0$, it can be seen that $\mathcal{D} > 0$ for all values of $\alpha$. 

\section{Direct-adjoint looping procedure}
\label{app:DAL}
When the forward equation is given by equation \eqref{eq:Ray_ft}, the adjoint equation is given by:
\begin{align}
    &\mathcal{F}^{\dagger}\hat{\xi} \equiv \partial_t(\partial_y^2 - \alpha^2)\hat{\xi} + i\alpha U(\partial_y^2 - \alpha^2)\hat{\xi} + 2i\alpha U'\partial_y\hat{\xi} = 0 \label{eq:Ray_adj_ft}
\end{align}
The adjoint variable $\hat{\xi}$ satisfies the same boundary conditions as $\hat{v}$.
  
At the outset, we define two inner products of functions $\hat{q}_1$ and $\hat{q}_2$ that will be useful for subsequent calculations:
\begin{align}
  &\langle \hat{q}_1,\hat{q}_2 \rangle = \langle \hat{q}_2,\hat{q}_1 \rangle = \frac{1}{2}\int^{y_u}_{y_l} \textrm{d}y~ \left(\hat{q}_2^*\hat{q}_1 + \hat{q}_1^* \hat{q}_2\right), \label{eq:spat_norm} \\
  &[\![ \hat{q}_1,\hat{q}_2 ]\!] = [\![ \hat{q}_2,\hat{q}_1 ]\!] = \frac{1}{2}\int^{T}_{0} \textrm{d}t~ \int^{y_u}_{y_l} \textrm{d}y~ \left(\hat{q}_2^* \hat{q}_1 + \hat{q}_1^*\hat{q}_2\right). \label{eq:spat_temp_norm}
\end{align}
If $\hat{q}_1$ and $\hat{q}_2$ that satisfy homogeneous boundary conditions at $y = y_l$ and $y = y_u$, it can be shown that:
\begin{align}
  \langle \hat{q}_1,\mathcal{M}\hat{q}_2 \rangle = \langle \mathcal{M}\hat{q}_1,\hat{q}_2 \rangle
\end{align}
The perturbation energy, using the form given in equation (\ref{eq:E}), can be written as:
\begin{align}
  E = \frac{1}{4\alpha^2}\langle \mathcal{M}\hat{v},\hat{v}\rangle
\end{align}
  
The objective functional for this study is the gain in perturbation energy at $t = T$ for a given set of base flow and perturbation parameters. This can be defined as follows:
\begin{align}
  \mathcal{J}(T) = \frac{E(T)}{E(0)} = \frac{\langle \mathcal{M}\hat{v}(T),\hat{v}(T)\rangle}{\langle \mathcal{M}\hat{v}(0),\hat{v}(0)\rangle} = \frac{\langle \mathcal{M}\hat{v}(T),\hat{v}(T)\rangle}{\langle \mathcal{M}\hat{v}_0,\hat{v}_0\rangle}
\end{align}
$\hat{v}_0$ is the initial condition, and we wish to find the initial condition that maximises the gain $\mathcal{J}$. It is also to be noted that the dependence on the spatial coordinate $y$ is not explicitly stated when it appears in the inner products of the form defined in equation (\ref{eq:spat_norm}). For the sake of brevity, only its temporal argument is specified explicitly.

We define the Lagrangian as follows:
\begin{align}
  &\mathcal{L} = \mathcal{J} - [\![\hat{\xi},\mathcal{F}\hat{v}]\!] - \langle \hat{g},\hat{v}(0) - \hat{v}_0\rangle \nonumber \\
  \Rightarrow \quad &\mathcal{L} = \frac{\langle \mathcal{M}\hat{v}(T),\hat{v}(T)\rangle}{\langle \mathcal{M}\hat{v}_0,\hat{v}_0\rangle} - \langle\hat{\xi}(T),\hat{v}(T)\rangle + \langle\hat{\xi}(0),\hat{v}(0)\rangle + [\![\mathcal{F}^{\dagger}\hat{\xi},\hat{v}]\!] - \langle \hat{g},\hat{v}(0) - \hat{v}_0\rangle
\end{align}
The adjoint state vector $\hat{\xi}$ and $\hat{g}$ serve as Lagrange multipliers.
  
To extremize $\mathcal{L}$ (and hence $\mathcal{J}$), we have to take variations of $\mathcal{L}$ with respect to different quantities. When the variations with respect to $\hat{\xi}$ and $\hat{g}$ are taken, the constraints are recovered (that is zero). In order to take variations with respect to $\hat{v}$ and $\hat{v}_0$, we define the following:
\begin{align}
  &\left[\frac{\partial \mathcal{L}}{\partial \hat{v}}, \delta \hat{v} \right] \equiv \lim_{\epsilon\to 0} \frac{\mathcal{L}\left(\hat{v} + \epsilon\delta\hat{v}\right) - \mathcal{L}\left(\hat{v}\right)}{\epsilon} \\
  &\left\langle \frac{\partial \mathcal{L}}{\partial \hat{v}_0}, \delta \hat{v}_0 \right\rangle \equiv \lim_{\epsilon\to 0} \frac{\mathcal{L}\left(\hat{v}_0 + \epsilon\delta\hat{v}_0\right) - \mathcal{L}\left(\hat{v}_0\right)}{\epsilon}
\end{align}
After taking variations with respect to $\hat{v}$ and $\hat{v}_0$, the conditions for extrema are: 
\begin{align}
  &\hat{\xi}(T) = \frac{2}{\langle\mathcal{M}\hat{v}_0,\hat{v}_0\rangle}\mathcal{M}\hat{v}(T), \label{eq:optcond_q}\\
  &\hat{\xi}(0) = \hat{g} = \frac{2\langle\mathcal{M}\hat{v}(T),\hat{v}(T)\rangle}{\langle\mathcal{M}\hat{v}_0,\hat{v}_0\rangle^2}\mathcal{M}\hat{v}_0. \label{eq:optcond_q0}
\end{align}

We attempt to find the initial condition $\hat{v}_0$ that gives an extrema for $\mathcal{J}$ at a specified target time $T$ in an iterative manner. With an initial guess for $\hat{v}_0$, $\hat{v}(T)$ is obtained by marching forward in time the direct equation \eqref{eq:Ray_ft}. $\hat{\xi}(T)$ can then be defined using equation (\ref{eq:optcond_q}). In the next step, $\hat{\xi}(0)$ by evolving $\hat{\xi}(T)$ backward in time using equation (\ref{eq:Ray_adj_ft}). The initial condition $\hat{v}_0$ can be then updated using equation \eqref{eq:optcond_q0}. The steps are repeated until criteria for terminating the iterations are satisfied. The ideal criterion for stopping the iterative procedure is when \eqref{eq:optcond_q0} is satisfied (up to a specified tolerance); this implies a global extrema has been found. However, it is not always guaranteed that this condition can be satisfied. Our stopping criteria is based on whether $\mathcal{J}$ undergoes significant change between iterations: the looping is terminated when the change in $\mathcal{J}$ in subsequent iterations is less than 0.1\%.

\bibliography{ref_nmod}

\begin{thebibliography}{58}%
\makeatletter
\providecommand \@ifxundefined [1]{%
 \@ifx{#1\undefined}
}%
\providecommand \@ifnum [1]{%
 \ifnum #1\expandafter \@firstoftwo
 \else \expandafter \@secondoftwo
 \fi
}%
\providecommand \@ifx [1]{%
 \ifx #1\expandafter \@firstoftwo
 \else \expandafter \@secondoftwo
 \fi
}%
\providecommand \natexlab [1]{#1}%
\providecommand \enquote  [1]{``#1''}%
\providecommand \bibnamefont  [1]{#1}%
\providecommand \bibfnamefont [1]{#1}%
\providecommand \citenamefont [1]{#1}%
\providecommand \href@noop [0]{\@secondoftwo}%
\providecommand \href [0]{\begingroup \@sanitize@url \@href}%
\providecommand \@href[1]{\@@startlink{#1}\@@href}%
\providecommand \@@href[1]{\endgroup#1\@@endlink}%
\providecommand \@sanitize@url [0]{\catcode `\\12\catcode `\$12\catcode
  `\&12\catcode `\#12\catcode `\^12\catcode `\_12\catcode `\%12\relax}%
\providecommand \@@startlink[1]{}%
\providecommand \@@endlink[0]{}%
\providecommand \url  [0]{\begingroup\@sanitize@url \@url }%
\providecommand \@url [1]{\endgroup\@href {#1}{\urlprefix }}%
\providecommand \urlprefix  [0]{URL }%
\providecommand \Eprint [0]{\href }%
\providecommand \doibase [0]{https://doi.org/}%
\providecommand \selectlanguage [0]{\@gobble}%
\providecommand \bibinfo  [0]{\@secondoftwo}%
\providecommand \bibfield  [0]{\@secondoftwo}%
\providecommand \translation [1]{[#1]}%
\providecommand \BibitemOpen [0]{}%
\providecommand \bibitemStop [0]{}%
\providecommand \bibitemNoStop [0]{.\EOS\space}%
\providecommand \EOS [0]{\spacefactor3000\relax}%
\providecommand \BibitemShut  [1]{\csname bibitem#1\endcsname}%
\let\auto@bib@innerbib\@empty
\bibitem [{\citenamefont {Schmid}\ and\ \citenamefont
  {Henningson}(2001)}]{Schmid_Henningson_2001book}%
  \BibitemOpen
  \bibfield  {author} {\bibinfo {author} {\bibfnamefont {P.~J.}\ \bibnamefont
  {Schmid}}\ and\ \bibinfo {author} {\bibfnamefont {D.~S.}\ \bibnamefont
  {Henningson}},\ }\href {https://doi.org/10.1007/978-1-4613-0185-1} {\emph
  {\bibinfo {title} {Stability and {T}ransition in {S}hear {F}lows}}}\
  (\bibinfo  {publisher} {Springer-Verlag, New York},\ \bibinfo {year}
  {2001})\BibitemShut {NoStop}%
\bibitem [{\citenamefont {Schmid}(2007)}]{Schmid_2007ARFM}%
  \BibitemOpen
  \bibfield  {author} {\bibinfo {author} {\bibfnamefont {P.~J.}\ \bibnamefont
  {Schmid}},\ }\bibfield  {title} {\bibinfo {title} {Nonmodal {S}tability
  {T}heory},\ }\href {https://doi.org/10.1146/annurev.fluid.38.050304.092139}
  {\bibfield  {journal} {\bibinfo  {journal} {Annu.~Rev.~Fluid Mech.}\ }\textbf
  {\bibinfo {volume} {39}},\ \bibinfo {pages} {129} (\bibinfo {year}
  {2007})}\BibitemShut {NoStop}%
\bibitem [{\citenamefont {Schmid}\ and\ \citenamefont
  {Brandt}(2014)}]{Schmid_Brandt_2014AMR}%
  \BibitemOpen
  \bibfield  {author} {\bibinfo {author} {\bibfnamefont {P.~J.}\ \bibnamefont
  {Schmid}}\ and\ \bibinfo {author} {\bibfnamefont {L.}~\bibnamefont
  {Brandt}},\ }\bibfield  {title} {\bibinfo {title} {Analysis of {F}luid
  {S}ystems: {S}tability, {R}eceptivity, {S}ensitivity. {L}ecture notes from
  the {FLOW}-{NORDITA} {S}ummer {S}chool on {A}dvanced {I}nstability {M}ethods
  for {C}omplex {F}lows, {S}tockholm, {S}weden, 2013},\ }\href
  {https://doi.org/10.1115/1.4026375} {\bibfield  {journal} {\bibinfo
  {journal} {Appl.~Mech.~Rev.}\ }\textbf {\bibinfo {volume} {66}},\ \bibinfo
  {pages} {024803} (\bibinfo {year} {2014})}\BibitemShut {NoStop}%
\bibitem [{\citenamefont {Butler}\ and\ \citenamefont
  {Farrell}(1992)}]{Butler_Farrell_1992PFA}%
  \BibitemOpen
  \bibfield  {author} {\bibinfo {author} {\bibfnamefont {K.~M.}\ \bibnamefont
  {Butler}}\ and\ \bibinfo {author} {\bibfnamefont {B.~F.}\ \bibnamefont
  {Farrell}},\ }\bibfield  {title} {\bibinfo {title} {Three-dimensional optimal
  perturbations in viscous shear flow},\ }\href
  {https://doi.org/10.1063/1.858386} {\bibfield  {journal} {\bibinfo  {journal}
  {Phys.~Fluids A}\ }\textbf {\bibinfo {volume} {4}},\ \bibinfo {pages} {1637}
  (\bibinfo {year} {1992})}\BibitemShut {NoStop}%
\bibitem [{\citenamefont {Trefethen}\ \emph {et~al.}(1993)\citenamefont
  {Trefethen}, \citenamefont {Trefethen}, \citenamefont {Reddy},\ and\
  \citenamefont {Driscoll}}]{Trefethen_etal_1993Science}%
  \BibitemOpen
  \bibfield  {author} {\bibinfo {author} {\bibfnamefont {L.~N.}\ \bibnamefont
  {Trefethen}}, \bibinfo {author} {\bibfnamefont {A.~E.}\ \bibnamefont
  {Trefethen}}, \bibinfo {author} {\bibfnamefont {S.~C.}\ \bibnamefont
  {Reddy}},\ and\ \bibinfo {author} {\bibfnamefont {T.~A.}\ \bibnamefont
  {Driscoll}},\ }\bibfield  {title} {\bibinfo {title} {Hydrodynamic {S}tability
  {W}ithout {E}igenvalues},\ }\href
  {https://doi.org/10.1126/science.261.5121.578} {\bibfield  {journal}
  {\bibinfo  {journal} {Science}\ }\textbf {\bibinfo {volume} {261}},\ \bibinfo
  {pages} {578} (\bibinfo {year} {1993})}\BibitemShut {NoStop}%
\bibitem [{\citenamefont {Bottin}\ \emph {et~al.}(1998)\citenamefont {Bottin},
  \citenamefont {Dauchot}, \citenamefont {Daviaud},\ and\ \citenamefont
  {Manneville}}]{Bottin_etal_1998PF}%
  \BibitemOpen
  \bibfield  {author} {\bibinfo {author} {\bibfnamefont {S.}~\bibnamefont
  {Bottin}}, \bibinfo {author} {\bibfnamefont {O.}~\bibnamefont {Dauchot}},
  \bibinfo {author} {\bibfnamefont {F.}~\bibnamefont {Daviaud}},\ and\ \bibinfo
  {author} {\bibfnamefont {P.}~\bibnamefont {Manneville}},\ }\bibfield  {title}
  {\bibinfo {title} {Experimental evidence of streamwise vortices as finite
  amplitude solutions in transitional plane {C}ouette flow},\ }\href
  {https://doi.org/10.1063/1.869773} {\bibfield  {journal} {\bibinfo  {journal}
  {Phys.~Fluids}\ }\textbf {\bibinfo {volume} {10}},\ \bibinfo {pages} {2597}
  (\bibinfo {year} {1998})}\BibitemShut {NoStop}%
\bibitem [{\citenamefont {Elofsson}\ \emph {et~al.}(1999)\citenamefont
  {Elofsson}, \citenamefont {Kawakami},\ and\ \citenamefont
  {Alfredsson}}]{Elofsson_etal_1999PF}%
  \BibitemOpen
  \bibfield  {author} {\bibinfo {author} {\bibfnamefont {P.~A.}\ \bibnamefont
  {Elofsson}}, \bibinfo {author} {\bibfnamefont {M.}~\bibnamefont {Kawakami}},\
  and\ \bibinfo {author} {\bibfnamefont {P.~H.}\ \bibnamefont {Alfredsson}},\
  }\bibfield  {title} {\bibinfo {title} {Experiments on the stability of
  streamwise streaks in plane {P}oiseuille flow},\ }\href
  {https://doi.org/10.1063/1.869962} {\bibfield  {journal} {\bibinfo  {journal}
  {Phys.~Fluids}\ }\textbf {\bibinfo {volume} {11}},\ \bibinfo {pages} {915}
  (\bibinfo {year} {1999})}\BibitemShut {NoStop}%
\bibitem [{\citenamefont {Andersson}\ \emph {et~al.}(2001)\citenamefont
  {Andersson}, \citenamefont {Brandt}, \citenamefont {Bottaro},\ and\
  \citenamefont {Henningson}}]{Andersson_etal_2001JFM}%
  \BibitemOpen
  \bibfield  {author} {\bibinfo {author} {\bibfnamefont {P.}~\bibnamefont
  {Andersson}}, \bibinfo {author} {\bibfnamefont {L.}~\bibnamefont {Brandt}},
  \bibinfo {author} {\bibfnamefont {A.}~\bibnamefont {Bottaro}},\ and\ \bibinfo
  {author} {\bibfnamefont {D.~S.}\ \bibnamefont {Henningson}},\ }\bibfield
  {title} {\bibinfo {title} {On the breakdown of boundary layer streaks},\
  }\href {https://doi.org/10.1017/S0022112000002421} {\bibfield  {journal}
  {\bibinfo  {journal} {J.~Fluid~Mech.}\ }\textbf {\bibinfo {volume} {428}},\
  \bibinfo {pages} {29} (\bibinfo {year} {2001})}\BibitemShut {NoStop}%
\bibitem [{\citenamefont {Schrader}\ \emph {et~al.}(2011)\citenamefont
  {Schrader}, \citenamefont {Brandt},\ and\ \citenamefont
  {Zaki}}]{Schrader_etal_11JFM}%
  \BibitemOpen
  \bibfield  {author} {\bibinfo {author} {\bibfnamefont {L.-U.}\ \bibnamefont
  {Schrader}}, \bibinfo {author} {\bibfnamefont {L.}~\bibnamefont {Brandt}},\
  and\ \bibinfo {author} {\bibfnamefont {T.~A.}\ \bibnamefont {Zaki}},\
  }\bibfield  {title} {\bibinfo {title} {Receptivity, instability and breakdown
  of {G}{\"o}rtler flow},\ }\href {https://doi.org/10.1017/jfm.2011.229}
  {\bibfield  {journal} {\bibinfo  {journal} {J.~Fluid~Mech.}\ }\textbf
  {\bibinfo {volume} {682}},\ \bibinfo {pages} {362} (\bibinfo {year}
  {2011})}\BibitemShut {NoStop}%
\bibitem [{\citenamefont {Lucas}\ \emph {et~al.}(2015)\citenamefont {Lucas},
  \citenamefont {Vermeersch},\ and\ \citenamefont
  {Arnal}}]{Lucas_etal_15EJMBF}%
  \BibitemOpen
  \bibfield  {author} {\bibinfo {author} {\bibfnamefont {J.-M.}\ \bibnamefont
  {Lucas}}, \bibinfo {author} {\bibfnamefont {O.}~\bibnamefont {Vermeersch}},\
  and\ \bibinfo {author} {\bibfnamefont {D.}~\bibnamefont {Arnal}},\ }\bibfield
   {title} {\bibinfo {title} {Transient growth of {G}{\"o}rtler vortices in
  two-dimensional compressible boundary layers. application to surface
  waviness},\ }\href {https://doi.org/10.1016/j.euromechflu.2014.11.005}
  {\bibfield  {journal} {\bibinfo  {journal} {Eur.~J.~Mech.~B-Fluids}\ }\textbf
  {\bibinfo {volume} {50}},\ \bibinfo {pages} {132} (\bibinfo {year}
  {2015})}\BibitemShut {NoStop}%
\bibitem [{\citenamefont {Jose}\ \emph {et~al.}(2017)\citenamefont {Jose},
  \citenamefont {Kuzhimparampil}, \citenamefont {Pier},\ and\ \citenamefont
  {Govindarajan}}]{Jose_etal_2017PRF}%
  \BibitemOpen
  \bibfield  {author} {\bibinfo {author} {\bibfnamefont {S.}~\bibnamefont
  {Jose}}, \bibinfo {author} {\bibfnamefont {V.}~\bibnamefont
  {Kuzhimparampil}}, \bibinfo {author} {\bibfnamefont {B.}~\bibnamefont
  {Pier}},\ and\ \bibinfo {author} {\bibfnamefont {R.}~\bibnamefont
  {Govindarajan}},\ }\bibfield  {title} {\bibinfo {title} {Algebraic
  disturbances and their consequences in rotating channel flow transition},\
  }\href {https://doi.org/10.1103/PhysRevFluids.2.083901} {\bibfield  {journal}
  {\bibinfo  {journal} {Phys.~Rev.~Fluids}\ }\textbf {\bibinfo {volume} {2}},\
  \bibinfo {pages} {083901} (\bibinfo {year} {2017})}\BibitemShut {NoStop}%
\bibitem [{\citenamefont {Jose}\ and\ \citenamefont
  {Govindarajan}(2020)}]{Jose_Govindarajan_2020PRSA}%
  \BibitemOpen
  \bibfield  {author} {\bibinfo {author} {\bibfnamefont {S.}~\bibnamefont
  {Jose}}\ and\ \bibinfo {author} {\bibfnamefont {R.}~\bibnamefont
  {Govindarajan}},\ }\bibfield  {title} {\bibinfo {title} {Non-normal origin of
  modal instabilities in rotating plane shear flows},\ }\href
  {https://doi.org/10.1098/rspa.2019.0550} {\bibfield  {journal} {\bibinfo
  {journal} {Proc.~R.~Soc.~A}\ }\textbf {\bibinfo {volume} {476}},\ \bibinfo
  {pages} {20190550} (\bibinfo {year} {2020})}\BibitemShut {NoStop}%
\bibitem [{\citenamefont {Landahl}(1980)}]{Landahl_1980JFM}%
  \BibitemOpen
  \bibfield  {author} {\bibinfo {author} {\bibfnamefont {M.~T.}\ \bibnamefont
  {Landahl}},\ }\bibfield  {title} {\bibinfo {title} {A note on an algebraic
  instability of inviscid parallel shear flows},\ }\href
  {https://doi.org/10.1017/S0022112080000122} {\bibfield  {journal} {\bibinfo
  {journal} {J.~Fluid Mech.}\ }\textbf {\bibinfo {volume} {98}},\ \bibinfo
  {pages} {243} (\bibinfo {year} {1980})}\BibitemShut {NoStop}%
\bibitem [{\citenamefont {Brandt}(2014)}]{Brandt_2014EJMBF}%
  \BibitemOpen
  \bibfield  {author} {\bibinfo {author} {\bibfnamefont {L.}~\bibnamefont
  {Brandt}},\ }\bibfield  {title} {\bibinfo {title} {The lift-up effect: {T}he
  linear mechanism behind transition and turbulence in shear flows},\ }\href
  {https://doi.org/10.1016/j.euromechflu.2014.03.005} {\bibfield  {journal}
  {\bibinfo  {journal} {Eur.~J.~Mech.~B~Fluids}\ }\textbf {\bibinfo {volume}
  {47}},\ \bibinfo {pages} {80} (\bibinfo {year} {2014})}\BibitemShut {NoStop}%
\bibitem [{\citenamefont {Orr}(1907)}]{Orr_1907PRIAA_alt}%
  \BibitemOpen
  \bibfield  {author} {\bibinfo {author} {\bibfnamefont {W.~M.}\ \bibnamefont
  {Orr}},\ }\bibfield  {title} {\bibinfo {title} {The {S}tability or
  {I}nstability of the {S}teady {M}otions of a {P}erfect {L}iquid and of a
  {V}iscous {L}iquid.},\ }\href@noop {} {\bibfield  {journal} {\bibinfo
  {journal} {Proc.~R.~Irish~Acad.~A.}\ }\textbf {\bibinfo {volume} {27}},\
  \bibinfo {pages} {9} (\bibinfo {year} {1907})}\BibitemShut {NoStop}%
\bibitem [{\citenamefont {Waleffe}(1997)}]{Waleffe_1997PF}%
  \BibitemOpen
  \bibfield  {author} {\bibinfo {author} {\bibfnamefont {F.}~\bibnamefont
  {Waleffe}},\ }\bibfield  {title} {\bibinfo {title} {On a self-sustaining
  process in shear flows},\ }\href {https://doi.org/10.1063/1.869185}
  {\bibfield  {journal} {\bibinfo  {journal} {Phys.~Fluids}\ }\textbf {\bibinfo
  {volume} {9}},\ \bibinfo {pages} {883} (\bibinfo {year} {1997})}\BibitemShut
  {NoStop}%
\bibitem [{\citenamefont {Jim{\'e}nez}\ and\ \citenamefont
  {Pinelli}(1999)}]{Jimenez_Pinelli_1999JFM}%
  \BibitemOpen
  \bibfield  {author} {\bibinfo {author} {\bibfnamefont {J.}~\bibnamefont
  {Jim{\'e}nez}}\ and\ \bibinfo {author} {\bibfnamefont {A.}~\bibnamefont
  {Pinelli}},\ }\bibfield  {title} {\bibinfo {title} {The autonomous cycle of
  near-wall turbulence},\ }\href {https://doi.org/10.1017/S0022112099005066}
  {\bibfield  {journal} {\bibinfo  {journal} {J.~Fluid~Mech.}\ }\textbf
  {\bibinfo {volume} {389}},\ \bibinfo {pages} {335} (\bibinfo {year}
  {1999})}\BibitemShut {NoStop}%
\bibitem [{\citenamefont {Panton}(2001)}]{Panton_2001PAS}%
  \BibitemOpen
  \bibfield  {author} {\bibinfo {author} {\bibfnamefont {R.~L.}\ \bibnamefont
  {Panton}},\ }\bibfield  {title} {\bibinfo {title} {Overview of the
  self-sustaining mechanisms of wall turbulence},\ }\href
  {https://doi.org/10.1016/S0376-0421(01)00009-4} {\bibfield  {journal}
  {\bibinfo  {journal} {Prog.~Aerosp.~Sci.}\ }\textbf {\bibinfo {volume}
  {37}},\ \bibinfo {pages} {341} (\bibinfo {year} {2001})}\BibitemShut
  {NoStop}%
\bibitem [{\citenamefont {Schoppa}\ and\ \citenamefont
  {Hussain}(2002)}]{Schoppa_Hussain_2002JFM}%
  \BibitemOpen
  \bibfield  {author} {\bibinfo {author} {\bibfnamefont {W.}~\bibnamefont
  {Schoppa}}\ and\ \bibinfo {author} {\bibfnamefont {F.}~\bibnamefont
  {Hussain}},\ }\bibfield  {title} {\bibinfo {title} {Coherent structure
  generation in near-wall turbulence},\ }\href
  {https://doi.org/10.1017/S002211200100667X} {\bibfield  {journal} {\bibinfo
  {journal} {J.~Fluid~Mech.}\ }\textbf {\bibinfo {volume} {453}},\ \bibinfo
  {pages} {57} (\bibinfo {year} {2002})}\BibitemShut {NoStop}%
\bibitem [{\citenamefont {Lozano-Dur{\'a}n}\ \emph {et~al.}(2021)\citenamefont
  {Lozano-Dur{\'a}n}, \citenamefont {Constantinou}, \citenamefont
  {Nikolaidis},\ and\ \citenamefont {Karp}}]{Lozano-Duran_etal_2021JFM}%
  \BibitemOpen
  \bibfield  {author} {\bibinfo {author} {\bibfnamefont {A.}~\bibnamefont
  {Lozano-Dur{\'a}n}}, \bibinfo {author} {\bibfnamefont {N.~C.}\ \bibnamefont
  {Constantinou}}, \bibinfo {author} {\bibfnamefont {M.-A.}\ \bibnamefont
  {Nikolaidis}},\ and\ \bibinfo {author} {\bibfnamefont {M.}~\bibnamefont
  {Karp}},\ }\bibfield  {title} {\bibinfo {title} {Cause-and-effect of linear
  mechanisms sustaining wall turbulence},\ }\href
  {https://doi.org/10.1017/jfm.2020.902} {\bibfield  {journal} {\bibinfo
  {journal} {J.~Fluid~Mech.}\ }\textbf {\bibinfo {volume} {914}},\ \bibinfo
  {pages} {A8} (\bibinfo {year} {2021})}\BibitemShut {NoStop}%
\bibitem [{\citenamefont {Jim{\'e}nez}(2013)}]{Jimenez_2013PF}%
  \BibitemOpen
  \bibfield  {author} {\bibinfo {author} {\bibfnamefont {J.}~\bibnamefont
  {Jim{\'e}nez}},\ }\bibfield  {title} {\bibinfo {title} {How linear is
  wall-bounded turbulence?},\ }\bibfield  {journal} {\bibinfo  {journal}
  {Phys.~Fluids}\ }\textbf {\bibinfo {volume} {25}},\ \href
  {https://doi.org/10.1017/jfm.2020.902} {10.1017/jfm.2020.902} (\bibinfo
  {year} {2013})\BibitemShut {NoStop}%
\bibitem [{\citenamefont {Jim{\'e}nez}(2018)}]{Jimenez_2018JFM}%
  \BibitemOpen
  \bibfield  {author} {\bibinfo {author} {\bibfnamefont {J.}~\bibnamefont
  {Jim{\'e}nez}},\ }\bibfield  {title} {\bibinfo {title} {Coherent structures
  in wall-bounded turbulence},\ }\href {https://doi.org/10.1017/jfm.2018.144}
  {\bibfield  {journal} {\bibinfo  {journal} {J.~Fluid~Mech.}\ }\textbf
  {\bibinfo {volume} {842}},\ \bibinfo {pages} {P1} (\bibinfo {year}
  {2018})}\BibitemShut {NoStop}%
\bibitem [{\citenamefont {Tissot}\ \emph {et~al.}(2017)\citenamefont {Tissot},
  \citenamefont {Laj{\'u}s~Jr.}, \citenamefont {Cavalieri},\ and\ \citenamefont
  {Jordan}}]{Tissot_etal_2017PRF}%
  \BibitemOpen
  \bibfield  {author} {\bibinfo {author} {\bibfnamefont {G.}~\bibnamefont
  {Tissot}}, \bibinfo {author} {\bibfnamefont {F.~C.}\ \bibnamefont
  {Laj{\'u}s~Jr.}}, \bibinfo {author} {\bibfnamefont {A.~V.~G.}\ \bibnamefont
  {Cavalieri}},\ and\ \bibinfo {author} {\bibfnamefont {P.}~\bibnamefont
  {Jordan}},\ }\bibfield  {title} {\bibinfo {title} {Wave packets and {O}rr
  mechanism in turbulent jets},\ }\href
  {https://doi.org/10.1103/PhysRevFluids.2.093901} {\bibfield  {journal}
  {\bibinfo  {journal} {Phys.~Rev.~Fluids}\ }\textbf {\bibinfo {volume} {2}},\
  \bibinfo {pages} {093901} (\bibinfo {year} {2017})}\BibitemShut {NoStop}%
\bibitem [{\citenamefont {Pickering}\ \emph {et~al.}(2020)\citenamefont
  {Pickering}, \citenamefont {Rigas}, \citenamefont {Nogueira}, \citenamefont
  {Cavalieri}, \citenamefont {Schmidt},\ and\ \citenamefont
  {Colonius}}]{Pickering_etal_2020JFM}%
  \BibitemOpen
  \bibfield  {author} {\bibinfo {author} {\bibfnamefont {E.}~\bibnamefont
  {Pickering}}, \bibinfo {author} {\bibfnamefont {G.}~\bibnamefont {Rigas}},
  \bibinfo {author} {\bibfnamefont {P.~A.~S.}\ \bibnamefont {Nogueira}},
  \bibinfo {author} {\bibfnamefont {A.~V.~G.}\ \bibnamefont {Cavalieri}},
  \bibinfo {author} {\bibfnamefont {O.~T.}\ \bibnamefont {Schmidt}},\ and\
  \bibinfo {author} {\bibfnamefont {T.}~\bibnamefont {Colonius}},\ }\bibfield
  {title} {\bibinfo {title} {{L}ift-up, {K}elvin--{H}elmholtz and {O}rr
  mechanisms in turbulent jets},\ }\href {https://doi.org/10.1017/jfm.2020.301}
  {\bibfield  {journal} {\bibinfo  {journal} {J.~Fluid~Mech.}\ }\textbf
  {\bibinfo {volume} {896}},\ \bibinfo {pages} {A2} (\bibinfo {year}
  {2020})}\BibitemShut {NoStop}%
\bibitem [{\citenamefont {Hack}\ and\ \citenamefont
  {Moin}(2017)}]{Hack_Moin_2017JFM}%
  \BibitemOpen
  \bibfield  {author} {\bibinfo {author} {\bibfnamefont {M.~J.~P.}\
  \bibnamefont {Hack}}\ and\ \bibinfo {author} {\bibfnamefont {P.}~\bibnamefont
  {Moin}},\ }\bibfield  {title} {\bibinfo {title} {Algebraic disturbance growth
  by interaction of {O}rr and lift-up mechanisms},\ }\href
  {https://doi.org/10.1017/jfm.2017.557} {\bibfield  {journal} {\bibinfo
  {journal} {J.~Fluid~Mech.}\ }\textbf {\bibinfo {volume} {829}},\ \bibinfo
  {pages} {112} (\bibinfo {year} {2017})}\BibitemShut {NoStop}%
\bibitem [{\citenamefont {Reddy}\ \emph {et~al.}(1998)\citenamefont {Reddy},
  \citenamefont {Schmid}, \citenamefont {Baggett},\ and\ \citenamefont
  {Henningson}}]{Reddy_etal_1998JFM}%
  \BibitemOpen
  \bibfield  {author} {\bibinfo {author} {\bibfnamefont {S.~C.}\ \bibnamefont
  {Reddy}}, \bibinfo {author} {\bibfnamefont {P.~J.}\ \bibnamefont {Schmid}},
  \bibinfo {author} {\bibfnamefont {J.~S.}\ \bibnamefont {Baggett}},\ and\
  \bibinfo {author} {\bibfnamefont {D.~S.}\ \bibnamefont {Henningson}},\
  }\bibfield  {title} {\bibinfo {title} {On stability of streamwise streaks and
  transition thresholds in plane channel flows},\ }\href
  {https://doi.org/10.1017/S0022112098001323} {\bibfield  {journal} {\bibinfo
  {journal} {J.~Fluid~Mech.}\ }\textbf {\bibinfo {volume} {365}},\ \bibinfo
  {pages} {269} (\bibinfo {year} {1998})}\BibitemShut {NoStop}%
\bibitem [{\citenamefont {Jiao}\ \emph {et~al.}(2021)\citenamefont {Jiao},
  \citenamefont {Hwang},\ and\ \citenamefont
  {Chernyshenko}}]{Jiao_etal_2021PRF}%
  \BibitemOpen
  \bibfield  {author} {\bibinfo {author} {\bibfnamefont {Y.}~\bibnamefont
  {Jiao}}, \bibinfo {author} {\bibfnamefont {Y.}~\bibnamefont {Hwang}},\ and\
  \bibinfo {author} {\bibfnamefont {S.~I.}\ \bibnamefont {Chernyshenko}},\
  }\bibfield  {title} {\bibinfo {title} {Orr mechanism in transition of
  parallel shear flow},\ }\href
  {https://doi.org/10.1103/PhysRevFluids.6.023902} {\bibfield  {journal}
  {\bibinfo  {journal} {Phys.~Rev.~Fluids}\ }\textbf {\bibinfo {volume} {6}},\
  \bibinfo {pages} {023902} (\bibinfo {year} {2021})}\BibitemShut {NoStop}%
\bibitem [{\citenamefont {Roy}\ and\ \citenamefont
  {Govindarajan}(2010)}]{Roy_Govindarajan_2010BookChap}%
  \BibitemOpen
  \bibfield  {author} {\bibinfo {author} {\bibfnamefont {A.}~\bibnamefont
  {Roy}}\ and\ \bibinfo {author} {\bibfnamefont {R.}~\bibnamefont
  {Govindarajan}},\ }\bibfield  {title} {\bibinfo {title} {An {I}ntroduction to
  {H}ydrodynamic {S}tability},\ }in\ \href
  {https://doi.org/10.1007/978-1-4419-6494-6_6} {\emph {\bibinfo {booktitle}
  {Rheology of {C}omplex {F}luids}}},\ \bibinfo {editor} {edited by\ \bibinfo
  {editor} {\bibfnamefont {A.~P.}\ \bibnamefont {Deshpande}}, \bibinfo {editor}
  {\bibfnamefont {J.~M.}\ \bibnamefont {Krishnan}},\ and\ \bibinfo {editor}
  {\bibfnamefont {P.~B.~S.}\ \bibnamefont {Kumar}}}\ (\bibinfo  {publisher}
  {Springer},\ \bibinfo {year} {2010})\ pp.\ \bibinfo {pages}
  {131--147}\BibitemShut {NoStop}%
\bibitem [{\citenamefont {Govindarajan}\ and\ \citenamefont
  {Sahu}(2014)}]{Govindarajan_Sahu_2014ARFM}%
  \BibitemOpen
  \bibfield  {author} {\bibinfo {author} {\bibfnamefont {R.}~\bibnamefont
  {Govindarajan}}\ and\ \bibinfo {author} {\bibfnamefont {K.~C.}\ \bibnamefont
  {Sahu}},\ }\bibfield  {title} {\bibinfo {title} {Instabilities in
  {V}iscosity-{S}tratified {F}low},\ }\href
  {https://doi.org/10.1146/annurev-fluid-010313-141351} {\bibfield  {journal}
  {\bibinfo  {journal} {Annu.~Rev.~Fluid~Mech.}\ }\textbf {\bibinfo {volume}
  {46}},\ \bibinfo {pages} {331} (\bibinfo {year} {2014})}\BibitemShut
  {NoStop}%
\bibitem [{\citenamefont {Ellingsen}\ and\ \citenamefont
  {Palm}(1975)}]{Ellingsen_Palm_1975PF}%
  \BibitemOpen
  \bibfield  {author} {\bibinfo {author} {\bibfnamefont {T.}~\bibnamefont
  {Ellingsen}}\ and\ \bibinfo {author} {\bibfnamefont {E.}~\bibnamefont
  {Palm}},\ }\bibfield  {title} {\bibinfo {title} {Stability of linear flow.},\
  }\href {https://doi.org/10.1063/1.861156} {\bibfield  {journal} {\bibinfo
  {journal} {Phys.~Fluids}\ }\textbf {\bibinfo {volume} {18}},\ \bibinfo
  {pages} {487} (\bibinfo {year} {1975})}\BibitemShut {NoStop}%
\bibitem [{\citenamefont {Roy}\ and\ \citenamefont
  {Subramanian}(2014)}]{Roy_Subramanian_2014JFM}%
  \BibitemOpen
  \bibfield  {author} {\bibinfo {author} {\bibfnamefont {A.}~\bibnamefont
  {Roy}}\ and\ \bibinfo {author} {\bibfnamefont {G.}~\bibnamefont
  {Subramanian}},\ }\bibfield  {title} {\bibinfo {title} {An inviscid modal
  interpretation of the `lift-up' effect},\ }\href
  {https://doi.org/10.1017/jfm.2014.485} {\bibfield  {journal} {\bibinfo
  {journal} {J.~Fluid~Mech.}\ }\textbf {\bibinfo {volume} {757}},\ \bibinfo
  {pages} {82} (\bibinfo {year} {2014})}\BibitemShut {NoStop}%
\bibitem [{\citenamefont {Craik}(1985)}]{Craik_1985book}%
  \BibitemOpen
  \bibfield  {author} {\bibinfo {author} {\bibfnamefont {A.~D.~D.}\
  \bibnamefont {Craik}},\ }\href {https://doi.org/10.1017/CBO9780511569548}
  {\emph {\bibinfo {title} {Wave interactions and fluid flows}}}\ (\bibinfo
  {publisher} {Cambridge University Press},\ \bibinfo {year}
  {1985})\BibitemShut {NoStop}%
\bibitem [{\citenamefont {Smyth}\ and\ \citenamefont
  {Carpenter}(2019)}]{Smyth_Carpenter_2019book}%
  \BibitemOpen
  \bibfield  {author} {\bibinfo {author} {\bibfnamefont {W.~D.}\ \bibnamefont
  {Smyth}}\ and\ \bibinfo {author} {\bibfnamefont {J.~R.}\ \bibnamefont
  {Carpenter}},\ }\href {https://doi.org/10.1017/9781108640084} {\emph
  {\bibinfo {title} {Instability in geophysical flows}}}\ (\bibinfo
  {publisher} {Cambridge University Press},\ \bibinfo {year}
  {2019})\BibitemShut {NoStop}%
\bibitem [{\citenamefont {Yih}(1967)}]{Yih_1967JFM}%
  \BibitemOpen
  \bibfield  {author} {\bibinfo {author} {\bibfnamefont {C.-S.}\ \bibnamefont
  {Yih}},\ }\bibfield  {title} {\bibinfo {title} {Instability due to viscosity
  stratification},\ }\href {https://doi.org/10.1017/S0022112067000357}
  {\bibfield  {journal} {\bibinfo  {journal} {J.~Fluid~Mech.}\ }\textbf
  {\bibinfo {volume} {27}},\ \bibinfo {pages} {337} (\bibinfo {year}
  {1967})}\BibitemShut {NoStop}%
\bibitem [{\citenamefont {Drazin}\ and\ \citenamefont
  {Reid}(2004)}]{Drazin_Reid_2004book}%
  \BibitemOpen
  \bibfield  {author} {\bibinfo {author} {\bibfnamefont {P.~G.}\ \bibnamefont
  {Drazin}}\ and\ \bibinfo {author} {\bibfnamefont {W.~H.}\ \bibnamefont
  {Reid}},\ }\href {https://doi.org/10.1017/CBO9780511616938} {\emph {\bibinfo
  {title} {Hydrodynamic {S}tability}}}\ (\bibinfo  {publisher} {Cambridge
  {U}niversity {P}ress, Cambridge},\ \bibinfo {year} {2004})\BibitemShut
  {NoStop}%
\bibitem [{\citenamefont {Howard}(1961)}]{Howard_1961JFM}%
  \BibitemOpen
  \bibfield  {author} {\bibinfo {author} {\bibfnamefont {L.~N.}\ \bibnamefont
  {Howard}},\ }\bibfield  {title} {\bibinfo {title} {Note on a paper of {John
  W. Miles}},\ }\href {https://doi.org/10.1017/S0022112061000317} {\bibfield
  {journal} {\bibinfo  {journal} {J.~Fluid~Mech.}\ }\textbf {\bibinfo {volume}
  {10}},\ \bibinfo {pages} {509} (\bibinfo {year} {1961})}\BibitemShut
  {NoStop}%
\bibitem [{\citenamefont {Huerre}\ and\ \citenamefont
  {Rossi}(1998)}]{Huerre_Rossi_1998BookChap}%
  \BibitemOpen
  \bibfield  {author} {\bibinfo {author} {\bibfnamefont {P.}~\bibnamefont
  {Huerre}}\ and\ \bibinfo {author} {\bibfnamefont {M.}~\bibnamefont {Rossi}},\
  }\bibfield  {title} {\bibinfo {title} {{H}ydrodynamic instabilities in open
  flows},\ }in\ \href {https://doi.org/10.1017/CBO9780511524608.004} {\emph
  {\bibinfo {booktitle} {{Hydrodynamics and Nonlinear Instabilities}}}},\
  \bibinfo {editor} {edited by\ \bibinfo {editor} {\bibfnamefont
  {C.}~\bibnamefont {Godr\`{e}che}}\ and\ \bibinfo {editor} {\bibfnamefont
  {P.}~\bibnamefont {Manneville}}}\ (\bibinfo  {publisher} {Cambridge
  University Press},\ \bibinfo {year} {1998})\ pp.\ \bibinfo {pages}
  {81--294}\BibitemShut {NoStop}%
\bibitem [{\citenamefont {Carpenter}\ \emph {et~al.}(2012)\citenamefont
  {Carpenter}, \citenamefont {Tedford}, \citenamefont {Heifetz},\ and\
  \citenamefont {Lawrence}}]{Carpenter_etal_2012AMR}%
  \BibitemOpen
  \bibfield  {author} {\bibinfo {author} {\bibfnamefont {J.~R.}\ \bibnamefont
  {Carpenter}}, \bibinfo {author} {\bibfnamefont {E.~W.}\ \bibnamefont
  {Tedford}}, \bibinfo {author} {\bibfnamefont {E.}~\bibnamefont {Heifetz}},\
  and\ \bibinfo {author} {\bibfnamefont {G.~A.}\ \bibnamefont {Lawrence}},\
  }\bibfield  {title} {\bibinfo {title} {Instability in {S}tratified {S}hear
  {F}low: Review of a {P}hysical {I}nterpretation {B}ased on {I}nteracting
  {W}aves},\ }\href {https://doi.org/10.1115/1.4007909} {\bibfield  {journal}
  {\bibinfo  {journal} {Appl.~Mech.~Rev.}\ }\textbf {\bibinfo {volume} {64}},\
  \bibinfo {pages} {060801} (\bibinfo {year} {2012})}\BibitemShut {NoStop}%
\bibitem [{\citenamefont {Case}(1960)}]{Case_1960PF}%
  \BibitemOpen
  \bibfield  {author} {\bibinfo {author} {\bibfnamefont {K.~M.}\ \bibnamefont
  {Case}},\ }\bibfield  {title} {\bibinfo {title} {Stability of {I}nviscid
  {P}lane {C}ouette {F}low},\ }\href {https://doi.org/10.1063/1.1706010}
  {\bibfield  {journal} {\bibinfo  {journal} {Phys.~Fluids}\ }\textbf {\bibinfo
  {volume} {3}},\ \bibinfo {pages} {143} (\bibinfo {year} {1960})}\BibitemShut
  {NoStop}%
\bibitem [{\citenamefont {Li}(1969)}]{Li_1969PF}%
  \BibitemOpen
  \bibfield  {author} {\bibinfo {author} {\bibfnamefont {C.-H.}\ \bibnamefont
  {Li}},\ }\bibfield  {title} {\bibinfo {title} {{Instability of Three-Layer
  Viscous Stratified Flow}},\ }\href {https://doi.org/10.1063/1.1692383}
  {\bibfield  {journal} {\bibinfo  {journal} {Phys.~Fluids}\ }\textbf {\bibinfo
  {volume} {12}},\ \bibinfo {pages} {2473} (\bibinfo {year}
  {1969})}\BibitemShut {NoStop}%
\bibitem [{\citenamefont {Weinstein}\ and\ \citenamefont
  {Kurz}(1991)}]{Weinstein_Kurz_1991PFA}%
  \BibitemOpen
  \bibfield  {author} {\bibinfo {author} {\bibfnamefont {S.~J.}\ \bibnamefont
  {Weinstein}}\ and\ \bibinfo {author} {\bibfnamefont {M.~R.}\ \bibnamefont
  {Kurz}},\ }\bibfield  {title} {\bibinfo {title} {Long-wavelength
  instabilities in three-layer flow down an incline},\ }\href
  {https://doi.org/10.1063/1.858158} {\bibfield  {journal} {\bibinfo  {journal}
  {Phys.~Fluids~A}\ }\textbf {\bibinfo {volume} {3}},\ \bibinfo {pages} {2680}
  (\bibinfo {year} {1991})}\BibitemShut {NoStop}%
\bibitem [{\citenamefont {Luchini}\ and\ \citenamefont
  {Bottaro}(2014)}]{Luchini_Bottaro_2014ARFM}%
  \BibitemOpen
  \bibfield  {author} {\bibinfo {author} {\bibfnamefont {P.}~\bibnamefont
  {Luchini}}\ and\ \bibinfo {author} {\bibfnamefont {A.}~\bibnamefont
  {Bottaro}},\ }\bibfield  {title} {\bibinfo {title} {Adjoint {E}quations in
  {S}tability {A}nalysis},\ }\href
  {https://doi.org/10.1146/annurev-fluid-010313-141253} {\bibfield  {journal}
  {\bibinfo  {journal} {Annu.~Rev.~Fluid~Mech.}\ }\textbf {\bibinfo {volume}
  {46}},\ \bibinfo {pages} {493} (\bibinfo {year} {2014})}\BibitemShut
  {NoStop}%
\bibitem [{\citenamefont {Jose}\ \emph {et~al.}(2020)\citenamefont {Jose},
  \citenamefont {Brandt},\ and\ \citenamefont
  {Govindarajan}}]{Jose_etal_2020IJHFF}%
  \BibitemOpen
  \bibfield  {author} {\bibinfo {author} {\bibfnamefont {S.}~\bibnamefont
  {Jose}}, \bibinfo {author} {\bibfnamefont {L.}~\bibnamefont {Brandt}},\ and\
  \bibinfo {author} {\bibfnamefont {R.}~\bibnamefont {Govindarajan}},\
  }\bibfield  {title} {\bibinfo {title} {Localisation of optimal perturbations
  in variable viscosity channel flow},\ }\href
  {https://doi.org/10.1016/j.ijheatfluidflow.2020.108588} {\bibfield  {journal}
  {\bibinfo  {journal} {Int.~J.~Heat~Fluid~Flow}\ }\textbf {\bibinfo {volume}
  {85}},\ \bibinfo {pages} {108588} (\bibinfo {year} {2020})}\BibitemShut
  {NoStop}%
\bibitem [{\citenamefont {Thakur}\ \emph {et~al.}(2021)\citenamefont {Thakur},
  \citenamefont {Sharma},\ and\ \citenamefont
  {Govindarajan}}]{Thakur_etal_2021JFM}%
  \BibitemOpen
  \bibfield  {author} {\bibinfo {author} {\bibfnamefont {R.}~\bibnamefont
  {Thakur}}, \bibinfo {author} {\bibfnamefont {A.}~\bibnamefont {Sharma}},\
  and\ \bibinfo {author} {\bibfnamefont {R.}~\bibnamefont {Govindarajan}},\
  }\bibfield  {title} {\bibinfo {title} {Early evolution of optimal
  perturbations in a viscosity-stratified channel},\ }\bibfield  {journal}
  {\bibinfo  {journal} {J.~Fluid~Mech.}\ }\textbf {\bibinfo {volume} {914}},\
  \href {https://doi.org/10.1017/jfm.2020.1160} {10.1017/jfm.2020.1160}
  (\bibinfo {year} {2021})\BibitemShut {NoStop}%
\bibitem [{\citenamefont {Briggs}\ \emph {et~al.}(1970)\citenamefont {Briggs},
  \citenamefont {Daugherty},\ and\ \citenamefont {Levy}}]{Briggs_etal_1970PF}%
  \BibitemOpen
  \bibfield  {author} {\bibinfo {author} {\bibfnamefont {R.~J.}\ \bibnamefont
  {Briggs}}, \bibinfo {author} {\bibfnamefont {J.~D.}\ \bibnamefont
  {Daugherty}},\ and\ \bibinfo {author} {\bibfnamefont {R.~H.}\ \bibnamefont
  {Levy}},\ }\bibfield  {title} {\bibinfo {title} {{Role of Landau Damping in
  Crossed-Field Electron Beams and Inviscid Shear Flow}},\ }\href
  {https://doi.org/10.1063/1.1692936} {\bibfield  {journal} {\bibinfo
  {journal} {Phys.~Fluids}\ }\textbf {\bibinfo {volume} {13}},\ \bibinfo
  {pages} {421} (\bibinfo {year} {1970})}\BibitemShut {NoStop}%
\bibitem [{\citenamefont {Schecter}\ \emph {et~al.}(2000)\citenamefont
  {Schecter}, \citenamefont {Dubin}, \citenamefont {Cass}, \citenamefont
  {Driscoll}, \citenamefont {Lansky},\ and\ \citenamefont
  {O'Neil}}]{Schecter_etal_2000PF}%
  \BibitemOpen
  \bibfield  {author} {\bibinfo {author} {\bibfnamefont {D.~A.}\ \bibnamefont
  {Schecter}}, \bibinfo {author} {\bibfnamefont {D.~H.~E.}\ \bibnamefont
  {Dubin}}, \bibinfo {author} {\bibfnamefont {A.~C.}\ \bibnamefont {Cass}},
  \bibinfo {author} {\bibfnamefont {C.~F.}\ \bibnamefont {Driscoll}}, \bibinfo
  {author} {\bibfnamefont {I.~M.}\ \bibnamefont {Lansky}},\ and\ \bibinfo
  {author} {\bibfnamefont {T.~M.}\ \bibnamefont {O'Neil}},\ }\bibfield  {title}
  {\bibinfo {title} {Inviscid damping of asymmetries on a two-dimensional
  vortex},\ }\href {https://doi.org/10.1063/1.1289505} {\bibfield  {journal}
  {\bibinfo  {journal} {Phys.~Fluids}\ }\textbf {\bibinfo {volume} {12}},\
  \bibinfo {pages} {2397} (\bibinfo {year} {2000})}\BibitemShut {NoStop}%
\bibitem [{\citenamefont {Shrira}\ and\ \citenamefont
  {Sazonov}(2001)}]{Shrira_Sazonov_2001JFM}%
  \BibitemOpen
  \bibfield  {author} {\bibinfo {author} {\bibfnamefont {V.~I.}\ \bibnamefont
  {Shrira}}\ and\ \bibinfo {author} {\bibfnamefont {I.~A.}\ \bibnamefont
  {Sazonov}},\ }\bibfield  {title} {\bibinfo {title} {Quasi-modes in
  boundary-layer-type flows. {P}art 1. {I}nviscid two-dimensional spatially
  harmonic perturbations},\ }\href {https://doi.org/10.1017/S0022112001005742}
  {\bibfield  {journal} {\bibinfo  {journal} {J.~Fluid~Mech.}\ }\textbf
  {\bibinfo {volume} {446}},\ \bibinfo {pages} {133} (\bibinfo {year}
  {2001})}\BibitemShut {NoStop}%
\bibitem [{\citenamefont {Dixit}\ and\ \citenamefont
  {Govindarajan}(2011)}]{Dixit_Govindarajan_2011JFM}%
  \BibitemOpen
  \bibfield  {author} {\bibinfo {author} {\bibfnamefont {H.~N.}\ \bibnamefont
  {Dixit}}\ and\ \bibinfo {author} {\bibfnamefont {R.}~\bibnamefont
  {Govindarajan}},\ }\bibfield  {title} {\bibinfo {title} {Stability of a
  vortex in radial density stratification: role of wave interactions},\ }\href
  {https://doi.org/10.1017/jfm.2011.156} {\bibfield  {journal} {\bibinfo
  {journal} {J.~Fluid~Mech.}\ }\textbf {\bibinfo {volume} {679}},\ \bibinfo
  {pages} {582} (\bibinfo {year} {2011})}\BibitemShut {NoStop}%
\bibitem [{\citenamefont {Polyachenko}\ and\ \citenamefont
  {Shukhman}(2022)}]{Polyachenko_Shukhman_2022PF}%
  \BibitemOpen
  \bibfield  {author} {\bibinfo {author} {\bibfnamefont {E.}~\bibnamefont
  {Polyachenko}}\ and\ \bibinfo {author} {\bibfnamefont {I.}~\bibnamefont
  {Shukhman}},\ }\bibfield  {title} {\bibinfo {title} {Damped perturbations in
  inviscid shear flows: van {K}ampen modes and {L}andau damping},\ }\href
  {https://doi.org/10.1063/5.0094089} {\bibfield  {journal} {\bibinfo
  {journal} {Phys.~Fluids}\ }\textbf {\bibinfo {volume} {34}},\ \bibinfo
  {pages} {064108} (\bibinfo {year} {2022})}\BibitemShut {NoStop}%
\bibitem [{\citenamefont {Constantin}\ and\ \citenamefont
  {Johnson}(2019)}]{Constantin_Johnson_2019BLM}%
  \BibitemOpen
  \bibfield  {author} {\bibinfo {author} {\bibfnamefont {A.}~\bibnamefont
  {Constantin}}\ and\ \bibinfo {author} {\bibfnamefont {R.~S.}\ \bibnamefont
  {Johnson}},\ }\bibfield  {title} {\bibinfo {title} {{Atmospheric Ekman flows
  with Variable Eddy Viscosity}},\ }\href
  {https://doi.org/10.1007/s10546-018-0404-0} {\bibfield  {journal} {\bibinfo
  {journal} {Boundary-Layer~Meteorol.}\ }\textbf {\bibinfo {volume} {170}},\
  \bibinfo {pages} {395} (\bibinfo {year} {2019})}\BibitemShut {NoStop}%
\bibitem [{\citenamefont {Dritschel}\ \emph {et~al.}(2020)\citenamefont
  {Dritschel}, \citenamefont {Paldor},\ and\ \citenamefont
  {Constantin}}]{Dritschel_etal_2020OS}%
  \BibitemOpen
  \bibfield  {author} {\bibinfo {author} {\bibfnamefont {D.~G.}\ \bibnamefont
  {Dritschel}}, \bibinfo {author} {\bibfnamefont {N.}~\bibnamefont {Paldor}},\
  and\ \bibinfo {author} {\bibfnamefont {A.}~\bibnamefont {Constantin}},\
  }\bibfield  {title} {\bibinfo {title} {The {E}kman spiral for
  piecewise-uniform viscosity},\ }\href
  {https://doi.org/10.5194/os-16-1089-2020} {\bibfield  {journal} {\bibinfo
  {journal} {Ocean~Sci.}\ }\textbf {\bibinfo {volume} {16}},\ \bibinfo {pages}
  {1089} (\bibinfo {year} {2020})}\BibitemShut {NoStop}%
\bibitem [{\citenamefont {Jose}\ \emph {et~al.}(2015)\citenamefont {Jose},
  \citenamefont {Roy}, \citenamefont {Bale},\ and\ \citenamefont
  {Govindarajan}}]{Jose_etal_2015PRSA}%
  \BibitemOpen
  \bibfield  {author} {\bibinfo {author} {\bibfnamefont {S.}~\bibnamefont
  {Jose}}, \bibinfo {author} {\bibfnamefont {A.}~\bibnamefont {Roy}}, \bibinfo
  {author} {\bibfnamefont {R.}~\bibnamefont {Bale}},\ and\ \bibinfo {author}
  {\bibfnamefont {R.}~\bibnamefont {Govindarajan}},\ }\bibfield  {title}
  {\bibinfo {title} {Analytical solutions for algebraic growth of disturbances
  in a stably stratified shear flow},\ }\href
  {https://doi.org/10.1098/rspa.2015.0267} {\bibfield  {journal} {\bibinfo
  {journal} {Proc.~R.~Soc.~A}\ }\textbf {\bibinfo {volume} {471}},\ \bibinfo
  {pages} {20150267} (\bibinfo {year} {2015})}\BibitemShut {NoStop}%
\bibitem [{\citenamefont {Jose}\ \emph {et~al.}(2018)\citenamefont {Jose},
  \citenamefont {Roy}, \citenamefont {Bale}, \citenamefont {Iyer},\ and\
  \citenamefont {Govindarajan}}]{Jose_etal_2018FDR}%
  \BibitemOpen
  \bibfield  {author} {\bibinfo {author} {\bibfnamefont {S.}~\bibnamefont
  {Jose}}, \bibinfo {author} {\bibfnamefont {A.}~\bibnamefont {Roy}}, \bibinfo
  {author} {\bibfnamefont {R.}~\bibnamefont {Bale}}, \bibinfo {author}
  {\bibfnamefont {K.}~\bibnamefont {Iyer}},\ and\ \bibinfo {author}
  {\bibfnamefont {R.}~\bibnamefont {Govindarajan}},\ }\bibfield  {title}
  {\bibinfo {title} {Optimal energy growth in a stably stratified shear flow},\
  }\href {https://doi.org/10.1088/1873-7005/aa838e} {\bibfield  {journal}
  {\bibinfo  {journal} {Fluid~Dyn.~Res.}\ }\textbf {\bibinfo {volume} {50}},\
  \bibinfo {pages} {011421} (\bibinfo {year} {2018})}\BibitemShut {NoStop}%
\bibitem [{\citenamefont {Farrell}\ and\ \citenamefont
  {Ioannou}(1993)}]{Farrell_Ioannou_1993JAS}%
  \BibitemOpen
  \bibfield  {author} {\bibinfo {author} {\bibfnamefont {B.~F.}\ \bibnamefont
  {Farrell}}\ and\ \bibinfo {author} {\bibfnamefont {P.~J.}\ \bibnamefont
  {Ioannou}},\ }\bibfield  {title} {\bibinfo {title} {Transient {D}evelopment
  of {P}erturbations in {S}tratified {S}hear {F}low},\ }\href
  {https://doi.org/10.1175/1520-0469(1993)050<2201:TDOPIS>2.0.CO;2} {\bibfield
  {journal} {\bibinfo  {journal} {J.~Atmos.~Sci.}\ }\textbf {\bibinfo {volume}
  {50}},\ \bibinfo {pages} {2201} (\bibinfo {year} {1993})}\BibitemShut
  {NoStop}%
\bibitem [{\citenamefont {Kaminski}\ \emph {et~al.}(2014)\citenamefont
  {Kaminski}, \citenamefont {Caulfield},\ and\ \citenamefont
  {Taylor}}]{Kaminski_etal_2014JFMR}%
  \BibitemOpen
  \bibfield  {author} {\bibinfo {author} {\bibfnamefont {A.~K.}\ \bibnamefont
  {Kaminski}}, \bibinfo {author} {\bibfnamefont {C.~P.}\ \bibnamefont
  {Caulfield}},\ and\ \bibinfo {author} {\bibfnamefont {J.~R.}\ \bibnamefont
  {Taylor}},\ }\bibfield  {title} {\bibinfo {title} {Transient growth in
  strongly stratified shear layers},\ }\href
  {https://doi.org/10.1017/jfm.2014.552} {\bibfield  {journal} {\bibinfo
  {journal} {J.~Fluid~Mech.}\ }\textbf {\bibinfo {volume} {758}},\ \bibinfo
  {pages} {R4} (\bibinfo {year} {2014})}\BibitemShut {NoStop}%
\bibitem [{\citenamefont {Vinokur}(1983)}]{Vinokur_1983JCP}%
  \BibitemOpen
  \bibfield  {author} {\bibinfo {author} {\bibfnamefont {M.}~\bibnamefont
  {Vinokur}},\ }\bibfield  {title} {\bibinfo {title} {{On One-Dimensional
  Stretching Functions for Finite-Difference Calculations}},\ }\href
  {https://doi.org/10.1016/0021-9991(83)90065-7} {\bibfield  {journal}
  {\bibinfo  {journal} {J.~Comput.~Phys.}\ }\textbf {\bibinfo {volume} {50}},\
  \bibinfo {pages} {215} (\bibinfo {year} {1983})}\BibitemShut {NoStop}%
\bibitem [{\citenamefont {Raghav}\ \emph {et~al.}(2022)\citenamefont {Raghav},
  \citenamefont {Jose}, \citenamefont {Apte},\ and\ \citenamefont
  {Govindarajan}}]{Raghav_etal_2022DAO}%
  \BibitemOpen
  \bibfield  {author} {\bibinfo {author} {\bibfnamefont {M.~S.}\ \bibnamefont
  {Raghav}}, \bibinfo {author} {\bibfnamefont {S.}~\bibnamefont {Jose}},
  \bibinfo {author} {\bibfnamefont {A.}~\bibnamefont {Apte}},\ and\ \bibinfo
  {author} {\bibfnamefont {R.}~\bibnamefont {Govindarajan}},\ }\bibfield
  {title} {\bibinfo {title} {{Effects of equatorially-confined shear flow on
  {MRG and R}ossby waves}},\ }\href
  {https://doi.org/10.1016/j.dynatmoce.2022.101331} {\bibfield  {journal}
  {\bibinfo  {journal} {Dyn.~Atmos.~Oceans}\ }\textbf {\bibinfo {volume}
  {100}},\ \bibinfo {pages} {101331} (\bibinfo {year} {2022})}\BibitemShut
  {NoStop}%
\bibitem [{\citenamefont {Weideman}\ and\ \citenamefont
  {Reddy}(2000)}]{Weideman_Reddy_2000ACM}%
  \BibitemOpen
  \bibfield  {author} {\bibinfo {author} {\bibfnamefont {J.~A.~C.}\
  \bibnamefont {Weideman}}\ and\ \bibinfo {author} {\bibfnamefont {S.~C.}\
  \bibnamefont {Reddy}},\ }\bibfield  {title} {\bibinfo {title} {A {MATLAB}
  {D}ifferentiation {M}atrix {S}uite},\ }\href
  {https://doi.org/10.1145/365723.365727} {\bibfield  {journal} {\bibinfo
  {journal} {ACM~Trans.~Math.~Software}\ }\textbf {\bibinfo {volume} {26}},\
  \bibinfo {pages} {465} (\bibinfo {year} {2000})}\BibitemShut {NoStop}%
\end{thebibliography}%
\end{document}